\documentclass[twocolumn,twocolappendix]{aastex631}

\newcommand{\github}[1]{%
   \href{#1}{\includegraphics[height=10pt,keepaspectratio]{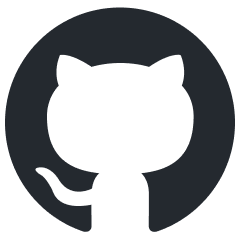}}
}

\usepackage{amssymb}
\usepackage{amsmath}
\usepackage[caption=false]{subfig}
\usepackage{booktabs}
\usepackage{verbatim}

\newcommand{\blockcomment}[1]{}
\usepackage{savesym}
\savesymbol{tablenum}
\usepackage{siunitx}
\restoresymbol{SIX}{tablenum}
\graphicspath{{./}{figures/}}
\begin{document}

\title{Mitigation of the Brighter-Fatter Effect in the LSST Camera}

\author[0000-0001-6966-5316]{Alex Broughton}
%\email{abrought@uci.edu}
\affiliation{Department of Physics and Astronomy, University of California-Irvine, Irvine, California, United States 92617}

\author[0000-0001-6161-8988]{Yousuke Utsumi}
%\email{youtsumi@stanford.edu}
\affiliation{Kavli Institute for Particle Astrophysics and Cosmology, Stanford University, Stanford, California, United States 94025}
\affiliation{SLAC National Accelerator Laboratory, Menlo Park, California, United States 94305-4085}

\author[0000-0002-2598-0514]{Andrés A. Plazas Malagón}
%\email{plazas@slac.stanford.edu}
\affiliation{Kavli Institute for Particle Astrophysics and Cosmology, Stanford University, Stanford, California, United States 94025}
\affiliation{SLAC National Accelerator Laboratory, Menlo Park, California, United States 94305-4085}

\author[0000-0003-1989-4879]{Christopher Waters}
%\email{czw@princeton.edu}
\affiliation{Department of Astrophysical Sciences, Princeton University, Princeton, NJ 08544, United States}

\author[0000-0002-9601-345X]{Craig Lage}
%\email{cslage@ucdavis.edu}
\affiliation{Department of Physics, University of California-Davis, Davis, California, United States 95616}

\author[0000-0002-2343-0949]{Adam Snyder}
%\email{aksnyder@ucdavis.edu}
\affiliation{Department of Physics, University of California-Davis, Davis, California, United States 95616}

\author[0000-0001-5738-8956]{Andrew Rasmussen}
%\email{arasmus@slac.stanford.edu}
\affiliation{Kavli Institute for Particle Astrophysics and Cosmology, Stanford University, Stanford, California, United States 94025}
\affiliation{SLAC National Accelerator Laboratory, Menlo Park, California, United States 94305-4085}

\author[0009-0001-0928-7445]{Stuart Marshall}
%\email{marshall@slac.stanford.edu}
\affiliation{Kavli Institute for Particle Astrophysics and Cosmology, Stanford University, Stanford, California, United States 94025}
\affiliation{SLAC National Accelerator Laboratory, Menlo Park, California, United States 94305-4085}

\author[0000-0001-5738-8956]{Jim Chiang}
%\email{jchiang@slac.stanford.edu}
\affiliation{Kavli Institute for Particle Astrophysics and Cosmology, Stanford University, Stanford, California, United States 94025}
\affiliation{SLAC National Accelerator Laboratory, Menlo Park, California, United States 94305-4085}

\author[0000-0003-0516-9422]{Simona Murgia}
%\email{smurgia@uci.edu}
\affiliation{Department of Physics and Astronomy, University of California-Irvine, Irvine, California, United States 92617}

\author[0000-0001-5326-3486]{Aaron Roodman}
%\email{roodman@slac.stanford.edu}
\affiliation{Kavli Institute for Particle Astrophysics and Cosmology, Stanford University, Stanford, California, United States 94025}
\affiliation{SLAC National Accelerator Laboratory, Menlo Park, California, United States 94305-4085}

%% automatically takes care of all commas and "and``*"s between authors names.


\begin{abstract}

Thick, fully depleted charge-coupled devices (CCDs) are known to exhibit non-linear behavior at high signal levels due to the dynamic behavior of charges collecting in the potential wells of pixels, called the brighter-fatter effect (BFE). This particularly impacts bright calibration stars, which appear larger than their intrinsic shape, creating a flux-dependent point-spread function (PSF) that if left unmitigated, could make up a large fraction of the error budget in Stage IV weak-lensing (WL) surveys such as the Legacy Survey of Space and Time (LSST). In this paper, we analyze image measurements of flat fields and artificial stars taken at different illumination levels with the LSST Camera (LSSTCam) at SLAC National Accelerator Laboratory in order to quantify this effect in the LSST Camera before and after a previously introduced correction technique. We observe that the BFE evolves anisotropically as a function of flux due to higher-order BFEs, which violates the fundamental assumption of this correction method. We then introduce a new sampling method based on a physically motivated model to account these higher-order terms in the correction, and then we test the modified correction on both datasets. We find that the new method corrects the effect in flat fields better than it corrects the effect in artificial stars which we conclude is the result of a unmodeled curl component of the deflection field by the correction.  We use these results to define a new metric for the full-well capacity of our sensors and advise image processing strategies to further limit the impact of the effect on LSST WL science pathways.

\end{abstract}

%% but this old ``*"keyword" functionality is maintained in case authors want


\section{Introduction} \label{sec:intro}

The Legacy Survey of Space and Time (LSST) will be conducted with the Simonyi Survey Telescope at the Vera C. Rubin Observatory, which is under construction on the summit of Cerro Pach\'{o}n in Chile. 
%The LSST Camera focal plane is made up of 189 individual charge-coupled devices (CCDs) designated for science use. It contains 3.2 gigapixels and will image a 9.6 deg$^{2}$ field of view with six filters (\textit{ugrizy}) covering the optical and near-IR parts of the spectrum. 
The survey plans to image 20,000 $\mathrm{deg}^2$ of the sky with $O(100)$, 15 second visits over 10 years across six filters (\textit{ugrizy}) in the optical and near-infrared (NIR) parts of the spectrum. It will map galaxies and optical transients to understand the natures of dark energy and dark matter and their impacts on the formation of structure in the universe \citep{lsst_SciDoc}. 

The instrument for this is a 3.2 gigapixel camera, which contains 201 individual charge-coupled devices (CCDs), with 189 designated specifically for science imaging. The sensors are fully depleted high-resistivity bulk silicon CCDs developed by two separate vendors. One type is made by Imaging Technology Laboratories (ITL), and the other is made by Teledyne e2v (E2V) to similar general architectural specifications. These sensors are arranged by type into $3\times3$ groups called raft-tower modules (RTMs) or rafts, which can each operate as an independent camera. Each CCD sensor is $4\,{\rm cm}\times4\,{\rm cm}$ made up of sixteen 1 megapixel channels each read out by its own amplifier and readout electronics. Each pixel on the LSST Camera focal plane is 10\,$\mu{\rm m} \times10\,\mu {\rm m}$ and has a depth of $100\,\mu {\rm m}$ \citep{raftpaper}.% and capable of $0.2$ arcsec sky-sampling. 
%The focal plane also contains 4 corner raft-tower modules (CRTMs), but these are used for exclusively for wavefront sensing and guiding and not scientific measurements. 

Much of the design and current development of instrumentation for LSST focuses on reducing the impact of systematic sensor artifacts in order to produce sub-percent level precision measurements of cosmological parameters and test currently prevailing thermodynamic models of the universe and theories of dark matter \citep{Albrecht2006}.

To meet these requirements, we need to understand the systematic effects in our sensors at the far limits of the capabilities of our instrumentation \citep{SciDriversToDesign}. Many of the problematic sensor behaviors are spatially static and can be simply calibrated or modeled as intrinsic constants at a particular location on a sensor and easily removed from raw images. However, this method would not work with locally variable or other signal-dependent effects, and such effects are known to exist \citep{Stubbs2013, Antilogus2014, Astier_2019}. 

Correlations between neighboring pixels have been shown to arise at high signal levels, at which point captured photocharges produce significant transverse electric fields on incoming photocharges \citep{Downing_2006,Holland2014, Lage_2017}. During the integration of an exposure, photo-electrons deflect into neighboring pixels in reaction to quasistatic changes in effective pixel area from the accumulated charges in the potential wells of the pixels, causing the measured light profile of a bright source to differ from that source's intrinsic surface brightness profile. This effect, dubbed the brighter-fatter effect (BFE), broadens intrinsic surface brightness profiles. The magnitude of the BFE depends on the surface brightness profile of the source itself and thus cannot be modeled solely by its location on a sensor. The consequence is that the BFE breaks the critical assumption of experimental imaging analysis that the pixels are independent light collectors that perfectly obey Poisson statistics. 

The BFE has been observed in LSSTCam sensors by \citet{Antilogus2014} and \citet{Lage_2017}, and in detectors used in other astronomical cameras such as Hyper Suprime-Cam (HSC) by \citet{Coulton_2018}, the Dark Energy Camera used by the Dark Energy Survey (DES) by \citet{gruen2015}, the Wide Field Camera 3 H1RG detector of the Hubble Space Telescope by \citet{plazas17}, MegaCam by \citet{Guyonnet2015}, the Mid-Infrared Instrument (MIRI) on board the James Webb Space Telescope (JWST) by \citet{jwstbf}, and in the near-infrared (NIR) detectors of the Wide Field Imager of NASA's Nancy Grace Roman Space Telescope \citep{Plazas_2018,hirata20,choi20,Freudenburg2020,Plazas_2023}. These studies measured a deviation from the Poissonian behavior of pixels in flat field images at bright illuminations, which they attributed to the BFE. Several corrections have been proposed by \citet{Antilogus2014}, \citet{gruen2015}, and \citet{Coulton_2018}, the last being a scalar algorithm which is currently used by in the HSC data reduction pipeline and is currently planned to be used in the LSST science pipelines \citep{HSCPipe, LsstDMPipeline1,LsstDMPipeline2}.

%The BFE has been measured by \citet{Lage_2017} in a single prototype LSST camera sensor made by Imaging Technologies Laboratory (ITL), using artificially produced point-sources, and they were able to accurately model the measured photometry to models of charge transport. 

In this paper, we will directly measure the BFE in the LSST Camera and our ability to correct it. In \S\ref{sec:bfcorrection}, we will introduce the theory behind the BFE and its correction by \citet{Coulton_2018} and weigh it in light of more recent findings by \citet{Astier_2019}. In \S\ref{sec:methods}, we will describe our laboratory measurements and image processing. In \S\ref{ssec:fullwellcapacitysection}--\ref{ssec:fittingthecovfunction}, we will explore the dynamic response of our sensors to charges in these data. In \S\ref{ssec:correcting_ptcs}--\ref{ssec:correcting_stars}, we will construct an improved application of the correction by \citet{Coulton_2018} and apply it to flat fields and artificial stars. In \S\ref{sec:discussion}, we will discuss the relative results between these two cases and directly test the underlying assumptions of the correction made by \citet{Coulton_2018}. 
Finally in \S\ref{ssec:bftocosmo}, we discuss how residual BFEs could influence several of LSST's dark energy science goals and motivate future work on analysis structures to mitigate its impact.

\section{Brighter-Fatter Correction}\label{sec:bfcorrection}
Several studies have shown that the pixel covariances in flat field images can be used to measure the BFE in the sensors of other optical and near-IR survey cameras \citep{Antilogus2014,gruen2015,Guyonnet2015,Coulton_2018, Astier_2023, hirata20, Freudenburg2020, plazas17, Plazas_2018}. The covariance can be numerically formalized between an arbitrary central pixel, $\mathbf{x} = (0, 0)$, assumed to be far from the edge of the detector, and a neighboring pixel $\mathbf{x'} = (i,j)$ in the difference of two flat field exposures, $F_1$ and $F_2$, with the same nominal signal level and  $\mu = \mathrm{avg}(F_1,F_2)$:
\begin{equation}
    \begin{split}
        C_{ij} &= \frac{1}{2}\mathrm{Cov}(F_1(\mathbf{x}) - F_2(\mathbf{x}), F_1(\mathbf{x'}) - F_2(\mathbf{x'})).
%        &= \mathrm{Cov}(F_1(\mathbf{x}), F_1(\mathbf{x'})) + \mathrm{Cov}(F_2(\mathbf{x}), F_2(\mathbf{x'})),
    \end{split}
\end{equation}

\noindent The directionality on each sensor is defined in regard to the CCDs' front-side architecture, where the vertical, $\mathbf{\hat{u}} = (0,1)$, direction is along the ``parallel" readout direction, and the horizontal, $\mathbf{\hat{u}} = (1, 0)$, direction is along the ``serial" readout direction.

\begin{figure}
    \centering
    \includegraphics[width=\linewidth]{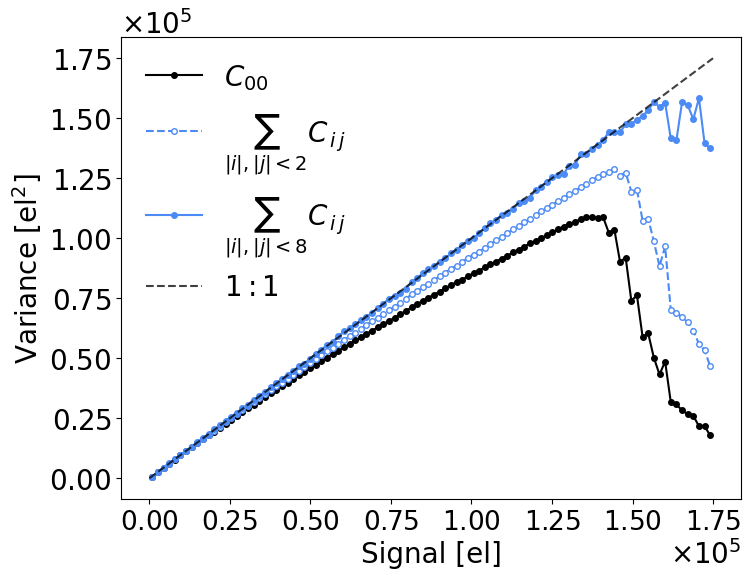}
    \caption{The variance and integrated covariance matrix vs. flux for a series of flat field images on an LSST camera sensor, averaging the covariance matrix for all amplifiers. We assume that the covariance matrix has parity symmetry about the central pixel, and we sum the covariance matrix from $|i,j| < 2$ and $|i,j| < 8$, and we show that summing out to 8 pixels fully reconstructs the Poissonian behavior. We used 342 total exposure pairs with $\times 1.025$ log-spacing in signal space, which were corrected with the basic instrument signature removal defined in Appendix \ref{appendix:configs}. We group the data into 100 bins from 0 to $1.75\times10^5\;el$. This was calculated for one sensor made by ITL on the LSST focal plane, which had an average gain of $1.702\;el\;/\;\mathrm{ADU}$ across all amplifiers. }
    \label{fig:summed_covariances}
\end{figure}

Figure \ref{fig:summed_covariances} shows an example of a densely sampled  measurement of the variance function $C_{00}(\mu)$ from a single sensor (ITL) on the LSST focal plane. If all the pixels are independent from each other, one can expect a simple linear relationship from the Poisson statistics as $C_{ij}(\mu)=(\delta_{i0}\delta_{j0})\mu/g +n_{ij}/g^2$, where $\delta_{ij}$ is Kronecker's delta, $g$ ($\mathrm{el}$/ADU) is the gain conversion between physical electrons and recorded counts (analog-digital units or ADU) by the analog-to-digital signal converter (ADC), and $n_{ij}$ [$\mathrm{e}^2$] is the noise at a pixel $(i,j)$. The measurement of $C_{00}$ departs from the linear relationship at high signal levels as excesses due to Poisson noise are suppressed due to the BFE since charges would be deflected out of pixels with excess accumulated charge and into neighboring pixels with less accumulated charge, resulting in a correlation between pixels at higher flat field signal levels. We measure a $30\%$ loss in the expected variance function (photon transfer curve or PTC) near pixel saturation (Figure~\ref{fig:summed_covariances}), which is consistent with the size of the effect observed by \citet{Astier_2019}, \citet{Coulton_2018}, \citet{Antilogus2014}, \citet{gruen2015}, \citet{Guyonnet2015}, \citet{Coulton_2018}, and \citet{Astier_2023}.

If charge is conserved, the sum of all the covariances away from this central pixel recovers all the pixel-level Poissonian sensor behavior lost due to the BFE \citep[first proposed by][]{Downing_2006}. The central pixel and the immediate neighbors should contribute the most to recovering the noise budget and integrating further away from the central pixel should add vanishingly smaller contributions. Figure \ref{fig:summed_covariances} also shows that accounting for the covariances 8px away from $(0, 0)$ in both directions on the focal plane allows us to reconstruct the Poissonian behavior in flat field exposures within statistical fluctuations up to fluxes above which the variance begins to drop off due to pixel saturation (around $1.35\times 10^5$ $\mathrm{el}$), for the two sensors that we tested.

\subsection{Modeling Pixel-Area Changes from Flat Field Statistics}\label{ssec:area_modeling}

We parameterize the BFE from flat-field statistics as changes in effective pixel area from distortions of a rectilinear grid, as proposed by \citet{Astier_2019}. We derive the pixel area distortions from the covariance function evaluated from flat fields. For covariances at a given signal level $\mu$, both expressed in ADU:

% FULLCOVARIANCE MODEL
\begin{equation}
    \begin{split}
    C_{ij}(\mu) = \frac{\mu}{g} \left [\delta_{i0}\delta_{j0} + a_{ij}\mu g + \frac{2}{3}[\mathbf{a}\otimes\mathbf{a} + \mathbf{ab}]_{ij}(\mu g)^{2} \right. \\
    \left. + \frac{1}{6}[2\mathbf{a}\otimes\mathbf{a}\otimes\mathbf{a} + 5\mathbf{a}\otimes\mathbf{ab}]_{ij}(\mu g)^{3} + \cdots \vphantom{\frac{1}{2}}\right ] + n_{ij}/g^2,
    \end{split}
    \label{eq:cov_model}
\end{equation}
where $a_{ij}$ $[1/e^{-}]$ describes the fractional change in pixel area due to changes in the pixel boundaries from accumulated source charges, $b_{ij}$ $[1/e^{-}]$  describes smaller time-dependent processes that weaken or strengthen the BFE as we will explore in a later section, $\mu$ is a given mean signal over the image region, $g$ is the sensor gain [${e}^- / \mathrm{ADU}$], $n_{ij}$ is a constant term that represents the mean square noise ($n_{00}\sim5.7e^{2}$, where the non-(0, 0) terms of the noise matrix are contributed by the electronics and overscan), $\mathbf{ab}$ is a direct matrix multiplication, and $\otimes$ refers to a discrete convolution operation.

This expansion allows for higher-order terms which are non-linear BFEs. \citet{Astier_2019} found this model to fit their data well, and the contribution from the non-linear terms was not negligible (with the $\mu^2$ term accounting for 20\% of the loss in variance). Any physically well-motivated correction algorithm would therefore need to take higher-order terms into account.

\subsection{Deriving a Scalar Correction with Higher-Order Effects}\label{ssec:ho_effects}

\citet{Coulton_2018} assumed that the deflection field along a single boundary is the gradient of a unitless, 2D scalar kernel $K$. In analogy with electrostatics, the effect of the kernel on the covariance matrix then takes the form of Poisson's equation with the Laplacian of the kernel \citep[equation 19 of ][]{Coulton_2018}:
\begin{equation}
    \widetilde{C}_{ij} = -\mu^2\nabla^2K_{ij}\label{eq:kernel}
\end{equation}
where
\begin{equation}
    \widetilde{C}_{ij}=C_{ij} - \left(\frac{\mu}{g} \delta_{i0}\delta_{j0}  + \frac{n_{ij}}{g^2}\right),
    \label{eq:cov_tilde}
\end{equation}

The true image can thus be recovered from the measured image using equation 22 of \citet{Coulton_2018}:
\begin{equation} \label{bfcorrectioneqn}
%    \begin{split}
%    &\hat{F}(\mathbf{x},t) = F(\mathbf{x},t) \\
%    &+ \frac{1}{2}\frac{\partial }{\partial x_i}\left ( F(\mathbf{x,t}) \frac{\partial }{\partial x_i}\int d^2\mathbf{x'}K(\mathbf{x}-\mathbf{x'})F(\mathbf{x'},t) \right ),
%    \end{split}
    \hat{F}(\mathbf{x}) = F(\mathbf{x}) + \frac{1}{2} \nabla \cdot \left[ F(\mathbf{x}) \nabla \left( K\otimes F(\mathbf{x})  \right)\right]
    \end{equation}
\noindent where $\hat{F}$ is the true image and $F$ is the measured image, and $\mathbf{x}$ runs over the pixel space in two dimensions. The expression includes a factor of $1/2$ to average the BFE over the duration of the exposure as the first charge to enter the pixel experiences no BFE and the last charge is assumed to experience twice the average effect.

The correction explicitly assumes that the covariance matrix does not evolve as a function of flux \citet{Coulton_2018}, which can be seen by the form of equation \ref{eq:cov_model}. However, \citet{Astier_2019} showed their functional form up to $O(\mu^4)$ fits the observed PTC well over the flux range they considered, citing $\chi^2 / \mathrm{N_{dof}} = 1.2$ for the fit to the variance element over all sensor channels in LSSTCam sensors, where they use $\mathrm{N_{dof}} = \mathrm{N(flat\;field\;pairs)} - 3$. This suggests that the PTC shape is contributed by other non-linear terms included in the covariance model. Whether to include the higher-order BFEs in the kernel, or to somehow weight the correction toward higher signal levels (where the BFE correction will be most important) becomes ambiguous, as there is no special signal level at which to sample $\widetilde{C}_{ij}$ and construct a kernel. Furthermore, the maximum charge capacity of CCDs are not well defined in practice or in literature. It then becomes important to measure the signal range over which the \citet{Coulton_2018} algorithm can reconstruct the impact of the BFE. % as compared to the tolerances required by the survey's science goals \citep{lsst_SciDoc}.

Correcting the BFE in flat-field images and linearizing the PTC allows us to directly test the impact of higher-order, flux-dependent effects. And correcting artificial star images allows us to test how well the \citet{Coulton_2018} correction can reconstruct shapes at all signal levels. Comparing the impact of the scalar correction on both flat-field images and star-field images is therefore a direct test of the validity of the underlying assumptions of the scalar correction.

\section{Data Acquisition and Image Processing}\label{sec:methods}
Our measurements of the BFE are derived from datasets taken during the fifth electro-optical testing period for the LSST Camera (informally referred to as Run 5), which took place on the integrated camera focal plane at the SLAC National Accelerator Laboratory in December 2021. In this section we describe the laboratory setup, data acquisition, and analysis methodology. This includes the configurations of the sensor testbed, acquisitions of calibration datasets and artificial star data, image processing pipeline, and photometry.

\subsection{Laboratory Setup}\label{labsetup}
We used the Bench for Optical Testing (BOT) as described in \citet{Newbry2018}. This test bench envelops the LSST Camera cryostat in a dark box that suppresses ambient light to a level $<0.01$ electrons per pixel per second and includes a rig to swap different optical projectors for illuminating the focal plane with various light patterns. This includes a \textit{spot grid projector}, which projects a $49\times49$ grid of stars (approximately $3\si{cm}\times3\si{cm}$ on the focal plane and approximately 65 pixels of rectilinear spacing between spots) to mimic a star field of various brightnesses, shapes, sizes, and positions (see Figure \ref{fig:spotexample}). The entire projector can be moved using remotely controlled motorized stages in all three axes for dithering (XY) across the focal plane and focusing (Z). This projector consists of a Nikon 105\,mm f/2.8 Al-s Micro-Nikkor lens, and an HTA Photomask photo-lithographic mask etched with pinholes. The spot grid pattern is illuminated by an integrating sphere with a 1" opening, which is fed by a 3\,mW fiber-coupled light emitting diode made by QPhotonics with a narrow band output peaked at 680\,nm. 

\subsection{Sensors Tested}

For this study of the BFE, we selected one top-performing sensor of each type, one ITL (R03-S12) and one E2V (R24-S11), to independently test the BF correction. The R-number refers to the raft row and column location on the focal plane and the S-number refers to the sensor row and column location within the raft. We selected these sensors based on serial Charge Transfer Inefficiency (sCTI) criterion \citep{2020arXiv200103223S}. Both sensors have measured sCTI for all channels up to $10^5\,e^{-}$, well below $10^{-6}$, which is within the needed requirements for LSST set in \citet{2016SPIE.9915E..0XO}. We also selected two other sensors, R02-S00 (ITL) and R22-S02 (E2V), which were used only as a secondary checks to compare broader performances between sensor types which we will explain in a later section, and the results we will show and discuss in this paper come from analyses of the two ``primary" sensors.

\subsection{Electro-Optical (EO) Datasets}\label{datacq}

\subsubsection{Calibrations}\label{ssec:calibdata}
We remove the static sensor effects from our lab exposures by subtracting calibrated images, which include 20 bias images to remove fixed-pattern noise, 20 dark images to eliminate thermal currents, and 342 densely sampled pairs of flat images from low flux (50 $\mathrm{e^-}/\mathrm{px}$) to high flux ($1.75\times10^5$ $\mathrm{e^-}/\mathrm{px}$) with a log-scale increment of $\times 1.025$ in the SDSS \textit{i}-band regime (700--800\,nm) to measure the covariance function, deferred charge, and linearity. For the flat pair acquisitions, we also have National Institute of Standards and Technology (NIST)-calibrated photo-diode measurements to provide an independent accurate measurement of the throughput of the light in the flat-field acquisition. 
%More detailed discussion of this point can be found in Appendix \ref{appendix:chromaticbf}.

We note that the SDSS \textit{i}-band (700\,nm--800\,nm), in which we measure our covariance function, covers a region of the electromagnetic spectrum outside the wavelength of the artificial star data (680\,nm). Chromatic dependence has not yet been rigorously studied in LSST sensors, however, based on preliminary studies by \citet{Astier_2023} in HSC, we believe any potential chromatic dependence of the BFE is negligible at these wavelengths that are comparatively close together. Further study of the BFE in the LSST Camera will investigate these hypothesized effects.

%We used this calibration dataset for subtracting 2D structure in bias, deffered charge correction, non-linearity correction, identifying and masking defects and deriving PTC.
%This dataset was used to remove other sources pixel-correlations, calibrate the BF correction, and characterize the dynamical properties of the sensor.

\subsubsection{Artificial Stars}\label{sssec:spot_data}
\begin{figure}[h!]
    \includegraphics[width=\linewidth]{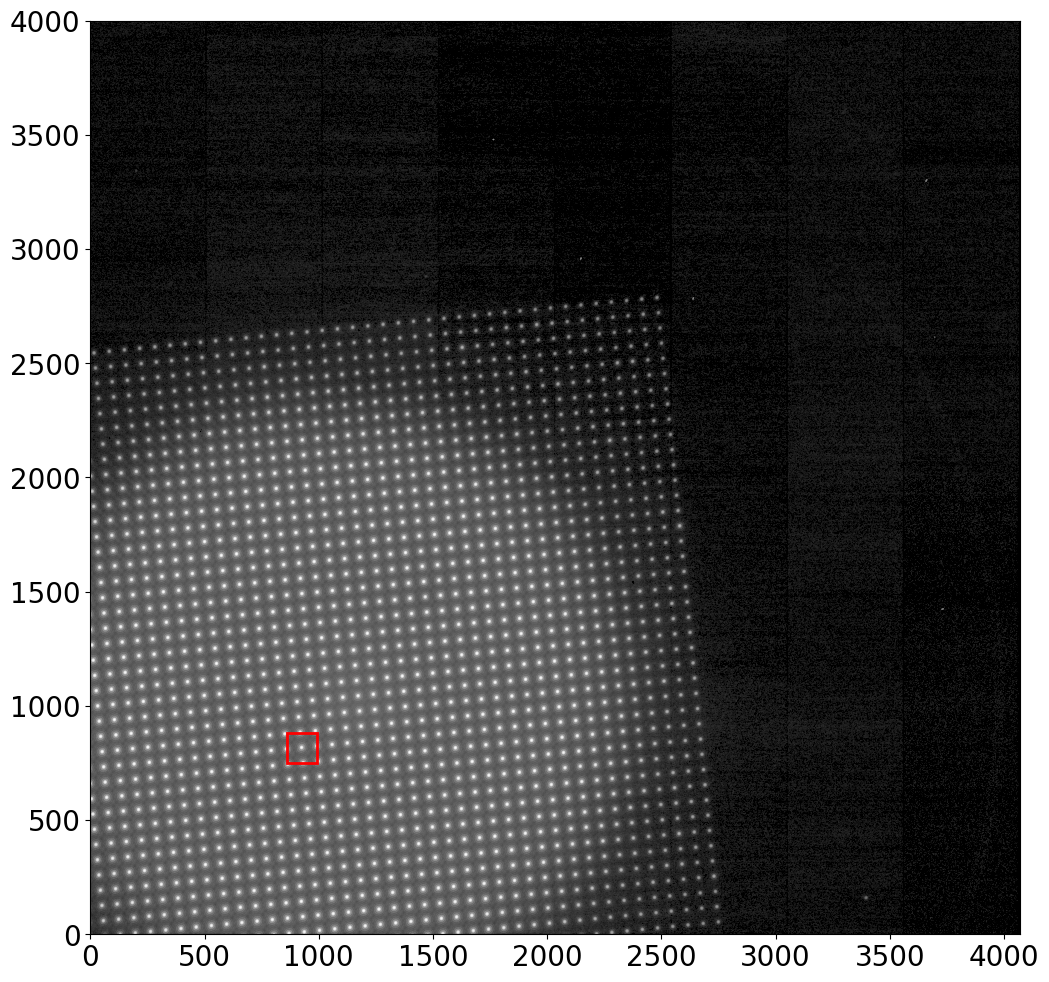}
    \begin{center}
        \hspace*{5mm}\includegraphics[scale=0.4]{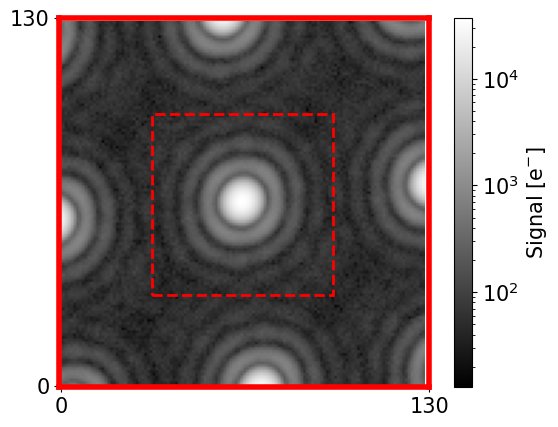}
    \end{center}
    \caption{Top: an example of a spot image (20\,s exposure with 680\,nm LED) that has been processed by standard instrument signature removal discussed in \S\ref{ssec:isr}. It shows the uneven (circularly symmetric) illumination pattern from the projector on a single sensor (ITL R03-S12). Bottom: an image subsection showing the Airy function systematic due to the long wavelength regime (solid red), and the corresponding $65\,\mathrm{px}\times65\,\mathrm{px}$ stamp (dashed-red) which sets the bounds used to derive calculate each star's shape.}
    %{This is an example of a spot image (10s exposure and 450nm LED) that has been processed by standard instrument signature removal (master bias subtracted). It was taken using the spot projector mounted to the Bench for Optical Testing and centered on a single sensor (E2V R22-S11).}
    \label{fig:spotexample}
\end{figure}
We projected artificial stars onto four sensors using the spot grid projector.
%The lens projects the $49\times49$ grid of artificial spots in a 65px$\times$65px rectilinear grid. The projection produces almost similar images of spots on the focal plane, with each spot approximately 4px FWHM at 680nm. 
An example exposure is shown in Figure \ref{fig:spotexample}. We project these spots onto the sensors during an exposure and fix the integrated light level by adjusting the integration time. We took a sequence of images of artificial stars at increasing intensities by varying the integration time in 15 steps from 5\,s to 200\,s to cover the full range in flux of our flat field calibrations (the full dynamic range of our CCDs), with 40 images at each exposure level. We split these image acquisitions in four quadrants around each sensor, changing the projector position 4 times across the whole imaging sequence. This allows us to measure many spots at a variety of positions and signal levels. We also ignore the first image after the projector changes position to avoid images with distorted stars that result from shaking the projector.  %These data were used to analyze the BFE in the low SNR regime and beyond the point of pixel saturation, where the pixel has filled up its full-well potential. 

%we find that the impact range of the systematic that we measure and the measurement of the shape statistics on a spot is dominated by the central few pixels, and there is negligible overlap between the diffraction rings of different spots. Therefore, we expect it's impact on shape-fitting measurements to also be negligible.

\subsection{Image Processing}
\subsubsection{Instrument Signature Removal (ISR)}\label{ssec:isr}

We processed each raw image using a complete sequence of the standard LSST Science Pipelines (w.2023.29 releases) and our calibrated Run 5 testing data products \citep{LsstDMPipeline1,LsstDMPipeline2}. The standard Instrument Signature Removal (ISR) removes  spatially characteristic systematic effects in the sensors, which includes bias subtraction with median-per-row overscan subtraction, dark current subtraction, non-linearity correction, and bad-pixel masking, and serial deferred charge. A detailed list of configurations used in the software are given in Appendix \ref{appendix:configs}.

We correct analog-to-digital converter (ADC) non-linearity \citep{Downing_2006} using a calibration correction informed by the densely sampled flat pairs compared with an independent measure of flux by the photodiode measurements. Additionally, we corrected the charge transfer inefficiency in the horizontal (serial) direction across the image using the same method as described in \citet{Snyder_2019} of \citet{Astier_2019}. We also applied bad-pixel masks and interpolated over the locations of bright and dark pixels and any cosmic-ray artifacts, though these pixels were excluded from calculations of the flat-field covariances.  We intentionally turned off the saturated pixel repair to avoid any non-physical distortions introduced by the repair. It is at this stage that we have the option to correct the BFE as well. The detailed set of options we chose for ISR is provided in Appendix 
\ref{appendix:configs}. 

We then apply the BFE correction itself. As described and recommended in the implementation of the correction \citep{Coulton_2018}, we recursively applied the kernel to each image until the total added charge in each step falls below a threshold of 10 electrons (the ``convergence condition"), which typically takes no more than 2 or 3 iterations.

\subsubsection{Photometry}

We measured the light profile and shape statistics of each artificial star from each exposure, deriving the source centroid, flux, and shape using the \texttt{Galsim} \citep{galsim} re-implementations of the Gaussian-weighted adaptive moments algorithm (HSM) defined by \citet{Hirata_2003} and tested independently by \citet{Mandelbaum_2005}. The HSM photometrics were derived for a square stamp around each star's centroid, with a side length equal to the spot grid spacing, which should entirely capture the star and any diffraction rings without presuming anything about the stars' shapes (see the bottom of Figure \ref{fig:spotexample}). The metrics include the diagonal elements of the quadrupole image moment tensor (here we denote them $I_{xx}$, $I_{yy}$, and $I_{xy}$) where $x$ is the serial direction along the orientation of the gates and $y$ is the parallel direction along the orientation of the channel stops. These statistics are useful because they are directly related to metrics of cosmic shear distortions. %More information about the calculation of these metrics can be found in Appendix \ref{appendix:imagemoments}.

\section{Results on Laboratory Acquisitions}\label{sec:results}

\subsection{Improved Benchmark Measurements of Full-Well Capacity}\label{ssec:fullwellcapacitysection}

\begin{figure*}[t]
    \newpage
    \gridline{\fig{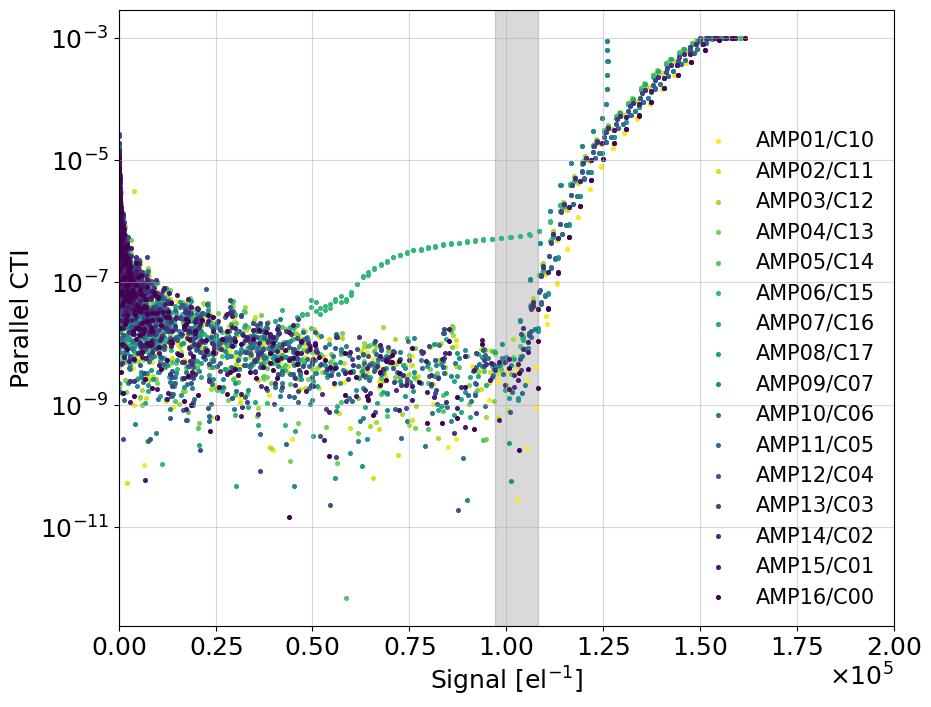}{0.49\textwidth}{(a) ITL Sensor (R03-S12)}
          \fig{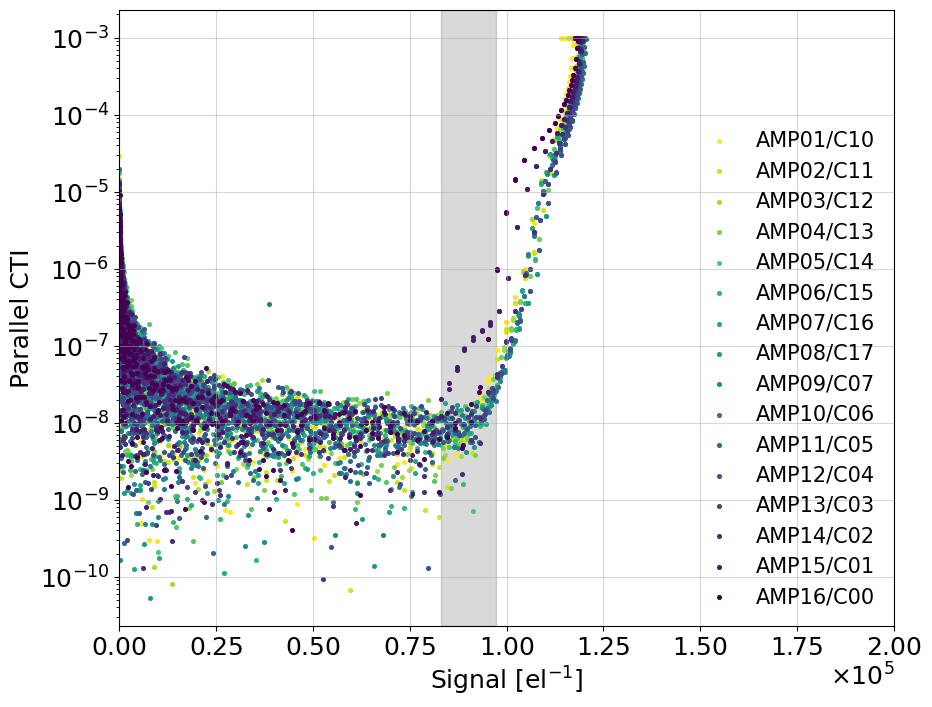}{0.49\textwidth}{(b) E2V Sensor (R24-S11)}}
    \caption{Measured parallel transfer CTI as functions of flux for the two sensors, with the range of turnoffs among the 16 channels highlighted. The anomalous amplifier in sensor R03-S12 (AMP06/C15) was ignored for all of our analyses due to abnormal pCTI behavior.}
    \label{fig:cti_turnoffs}
\end{figure*}

\begin{table*}
    \centering
    $C_{ij}(\mu)$ Fit Parameters
    \begin{tabular*}{\textwidth}{@{\extracolsep{\fill}}lcccc}
      \toprule[1pt] % <-- Toprule here
      \midrule[0.3pt]
      Parameter & R03-S12 (ITL) & R24-S11 (E2V) & R02-S00 (ITL) & R21-S02 (E2V) \\
      \midrule[0.3pt] % <-- Midrule here
      $a_{00}$  & $-1.80\times10^{-6}$ & $-3.11\times10^{-6}$ & $-1.80\times10^{-6}$ & $-3.10\times10^{-6}$ \\ 
      \midrule[0.3pt]
      $a_{10}$  & $1.18\times10^{-7}$ & $1.63\times10^{-7}$ & $1.32\times10^{-7}$ & $1.64\times10^{-7}$ \\ 
      \midrule[0.3pt]
      $a_{01}$  & $2.54\times10^{-7}$ & $3.83\times10^{-7}$ & $2.62\times10^{-7}$ & $3.78\times10^{-7}$ \\ 
      \midrule[0.3pt]
      $\sigma(a_{00})\;/\;|a_{00}|$  & $3.42\%$ & $3.44\%$ & $7.35\%$ & $3.94\%$ \\ 
      
      \midrule[0.3pt] % <-- Midrule here
      $b_{00}$  & $-8.69\times10^{-7}$ & $-2.95\times10^{-7}$ & $-8.10\times10^{-7}$ & $-5.38\times10^{-7}$ \\ 
      \midrule[0.3pt]
      $b_{10}$  & $-5.36\times10^{-6}$ & $-3.65\times10^{-6}$ & $-5.97\times10^{-6}$ & $-3.01\times10^{-6}$ \\ 
      \midrule[0.3pt]
      $b_{01}$  & $4.71\times10^{-8}$ & $2.04\times10^{-6}$ & $5.42\times10^{-7}$ & $1.06\times10^{-6}$ \\ 
      \midrule[0.3pt]
      $\sigma(b_{00})\;/\;|b_{00}|$  & $50.8\%$ & $171\%$ & $163\%$ & $133\%$ \\ 
      \midrule[0.3pt]
      $\bar{g}$ & 1.70 & 1.50 & 1.68 & 1.50 \\
      \bottomrule[0.3pt] % <-- Bottomrule here
    \end{tabular*}
    \caption{Fitted parameters of the full covariance model (equation \ref{eq:cov_model}) to calibrated data (\S \ref{ssec:calibdata}). We fit up to the pCTI turnoff, which was calibrated per amplifier, and then all parameters are averaged over all amplifiers for each sensor. Sigma parameters are given as the scatter over all amplifiers. }\label{tbl:cov_fit}
\end{table*}

We first assess the dynamic range of the CCDs from our calibration data.
We observe a correspondence between rapid changes of the covariance function with different metrics of pixel saturation or ``turnoffs" defined in previous literature, which we further correlate with physical mechanisms of charge transport. Typically, the pixel saturation level is defined as the level above which the flat images dramatically lose variance ($\mathrm{d}C_{00}/\mathrm{d}\mu < 0$) and no longer monotonically increases \citep[the ``PTC turnoff" as described by][]{Janesick2001}. Another method defined by \citet{Snyder_2019} assigns the limit as the flux above which charge can no longer be effectively transferred serially out of the sensor during readout. The serial charge transfer efficiency, sCTE \citep[or rather sCTI in the case of charge transfer in-efficiency described in][]{Rhodes2010} is calculated using the extended pixel edge response (EPER) method:
\begin{equation}
    \mathrm{CTI}(\mu) = \frac{F_{overscan}(\mu)}{N_{T}F_{lastpixel}(\mu)}
\end{equation}
\noindent where $F_{overscan} / F_{lastpixel}$ is the ratio between the total charge left in the overscan region after an image is transferred out to the last image column and $N_{T}$ is the number of pixel transfers during that readout, which is calculated for a specific flat image at a given flux level. The critical limit (``sCTI turnoff") is defined as the signal level at which the sCTI crosses above $10^{-5}$. Another method defines the maximum capacity more simply as the brightest recorded pixel (``maximum observed signal"). In our investigation, we discovered another relevant limit, which falls below all of these other limits. It can be identified as a turnoff in the parallel transfer component of CTI (``pCTI turnoff"), as shown in Figure \ref{fig:cti_turnoffs}, which is calculated analogously to the sCTI in \citet{Snyder_2019}. We define this point of pCTI turnoff by fitting the $\mathrm{pCTI}(\mu)$ to a flat line above the level of noise ($2.5\times10^4\;\mathrm{e^-}$), and below any other features ($5.0\times10^4\;\mathrm{e^-}$), and determine where the pCTI deviates by more than $3\sigma$ given by the standard error of the fitted data points. We typically measure this limit to be between $0.8-1.1\times10^5\;\mathrm{e}^{-}$, which is below any of our other metrics for CCD full-well capacity.

%Between the two CCD types, there are different underlying architectures of the gates and channel stops configurations as well as the bias and clocking voltages, but generally the confining electric fields are weaker in the parallel direction than they are in the serial direction. 

Each of these turnoff levels represents a different source of pixel correlations that begins to be relevant at different signal levels. And while we find the turnoff levels to vary from sensor to sensor, we find that their order is consistent in all the sensors we tested. The lowest is the pCTI turnoff ($0.83-1.13\times10^5\;\mathrm{e}^{-}$), followed by the PTC turnoff, the sCTI turnoff ($1.40-1.75\times10^5\;\mathrm{e}^{-}$), and the maximum observed signal ($>1.50\times10^5\;\mathrm{e}^{-}$). The different physical effects that determine these levels would bias the fit of our covariance model to the PTC, and our estimation of the strength of the BFE. We therefore limit the ranges of our fits to avoid including these other sources of pixel correlations that could be misattributed to the BFE. %, particularly below the point of pixel saturation. 

%There is some dependency between these turnoffs, and we see that when the correlation between the vertical pixels, there is a smaller loss in correlation with respect to the horizontal pixels since charge is conserved and charge that leaks in one direction is less likely to leak in the other direction.

%The covariance turnoff in the serial direction $C_{10}$ corresponds with the sCTI turnoff, and there is a second turnoff in the parallel direction $C_{01}$, which corresponds to the PTC turnoff. There is some dependency between these turnoffs, and we see that when the correlation between the vertical pixels, there is a smaller loss in correlation with respect to the horizontal pixels since charge is conserved and charge that leaks in one direction is less likely to leak in the other direction. While its effect on the full covariance function might be small and negligibly impact the electrostatic model fit, we do predict this to have a noticeable effect on our algorithm's ability to correct BF at high signal levels. In the course of our investigation, we considered the optimal signal domain on which to compute our model fit to the covariance function to derive the influence of the BFE. We ignore the flux regime above this point in our model fit to avoid misattributing other sources of pixel correlation to the shift in pixel areas due to BF alone. 
\begin{figure}
    \includegraphics[width=0.48\textwidth]{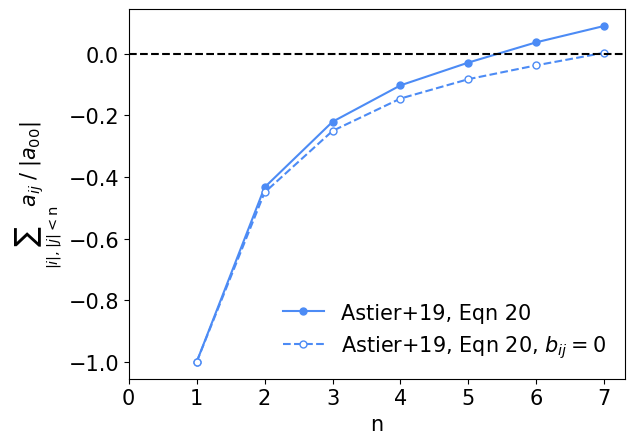}
    \caption{The charge conservation condition of the $a_{ij}$ model for the cases where $b_{ij}$ is allowed to vary or held fixed at 0. We calculate the sum over the $\mathbf{a}$ matrix.  This was calculated for the ITL sensor R03-S12.}
    \label{fig:charge_conservation_a_matrix}
\end{figure}

\begin{figure}
    \hfill\includegraphics[width=0.48\textwidth]{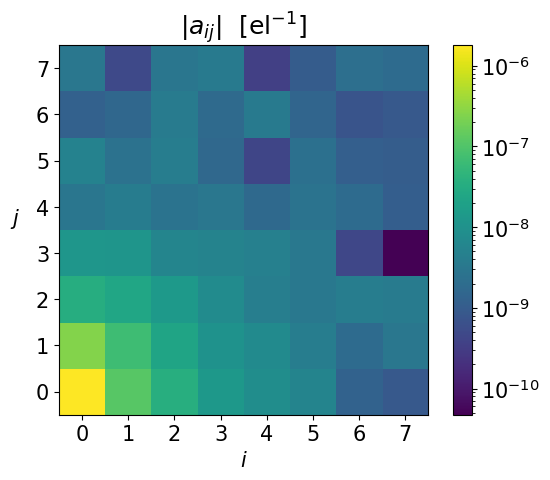}
    \includegraphics[width=0.48\textwidth]{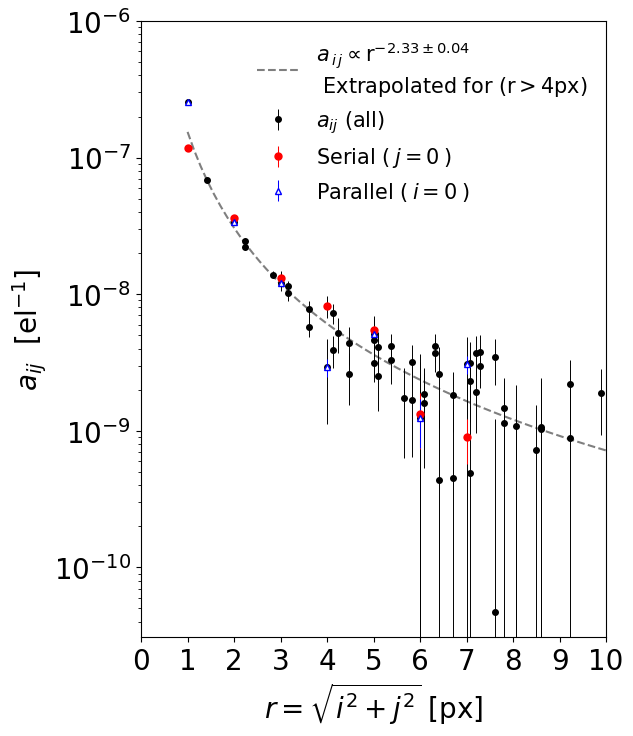}
    \caption{The pixel fractional area change as a matrix derived from the full covariance model (equation \ref{eq:cov_model}). The top panel shows the asymmetry in direction ($a_{01} / a_{10} \sim 2.16$), and the bottom panel shows the decay of the matrix into noise past $3-4$ pixels. We also show the serial and parallel pixels in red and blue to show the difference in the model in both directions on the sensor. The central term is $a_{00} = -1.80\times10^{-6}$. Errorbars are errors on the mean across all 16 amplifiers. This was calculated for an ITL sensor on the LSST focal plane (R03-S12).}
    \label{fig:amatrix}
\end{figure}

\begin{figure}
    \hfill\includegraphics[width=0.47\textwidth]{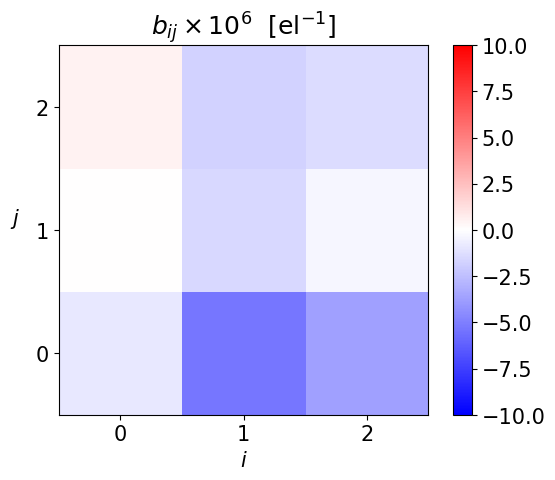}
    \includegraphics[width=0.49\textwidth]{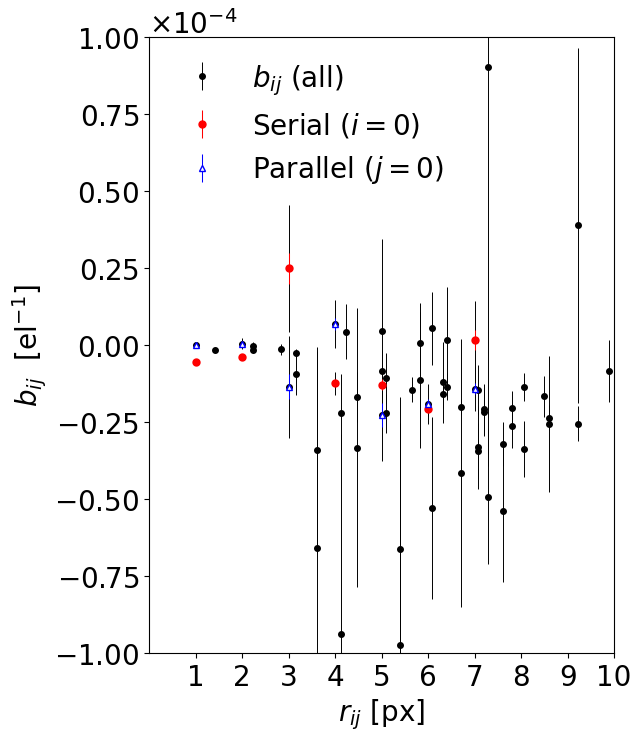}
    \caption{The measured $b_{ij}$ charge-displacement  component derived from the full covariance model (equation \ref{eq:cov_model}). We also show the serial and parallel pixels in red and blue to show the difference in the model in both directions on the sensor. It shows the relative increase (positive) or decrease (negative) shift in the accumulated charges' centroids. This was calculated for an ITL sensor on the LSST focal plane (R03-S12).}
    \label{fig:bmatrix}
\end{figure}

% template

\begin{figure*}[!p]
    \centering
    \gridline{\fig{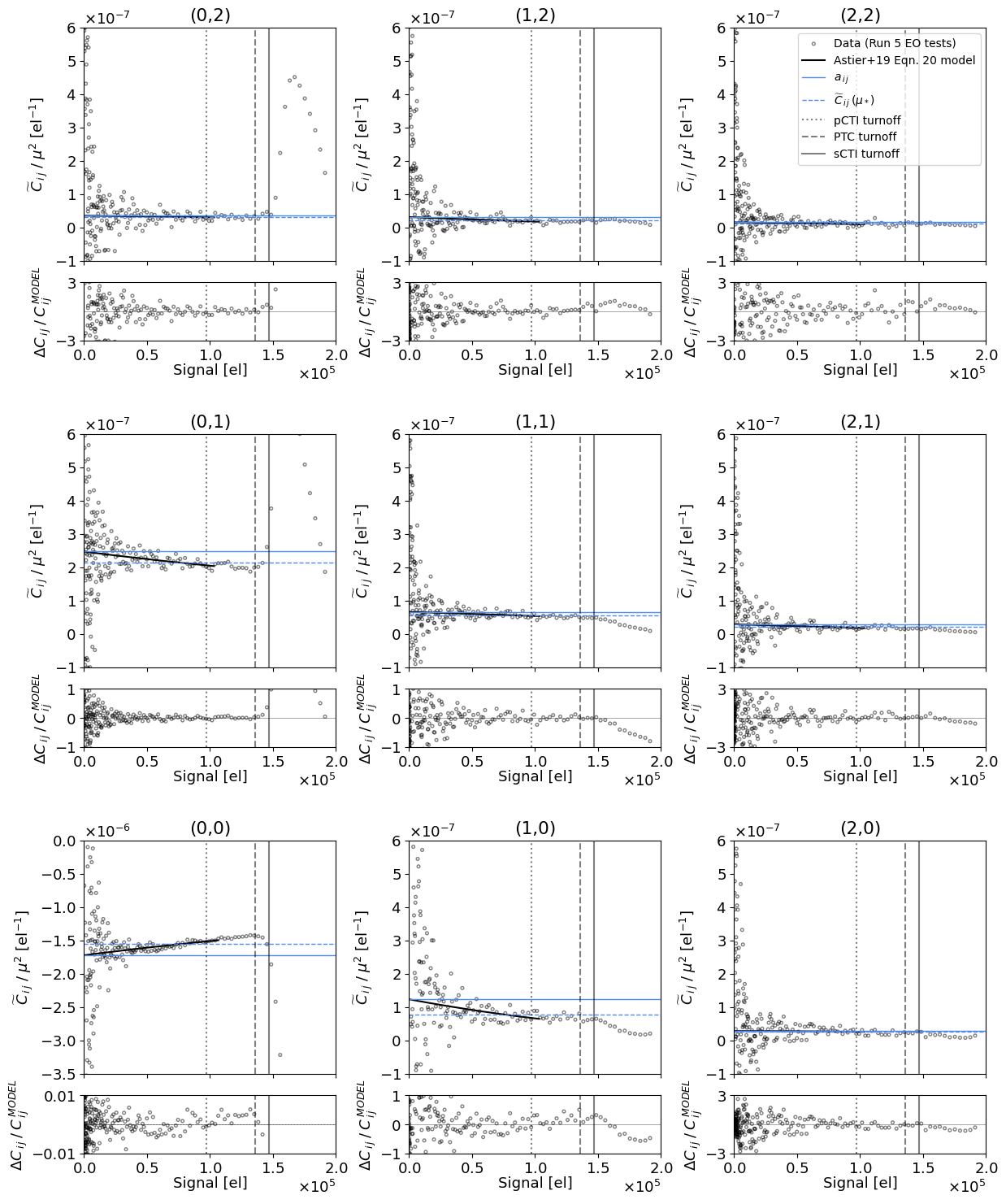}{0.95\linewidth}{(a) ITL Sensor (R03-S12)}}
    \caption{The higher-order term of the covariance for the ITL sensor (R03-S12). Top: the analytically calculated first-order component of the \citet{Astier_2019} covariance model, showing the increase in the fractional pixel area change with signal level and the corresponding fitted and scaled parameters for both sensor types. Bottom: the fractional residuals between the measured covariances and the corresponding covariance model. The vertical lines mark the various measures of full-well conditions.\label{fig:analytical_a_matrix_itl}}
\end{figure*}
\begin{figure*}[!p]\ContinuedFloat
    \medskip
    \centering
    \gridline{\fig{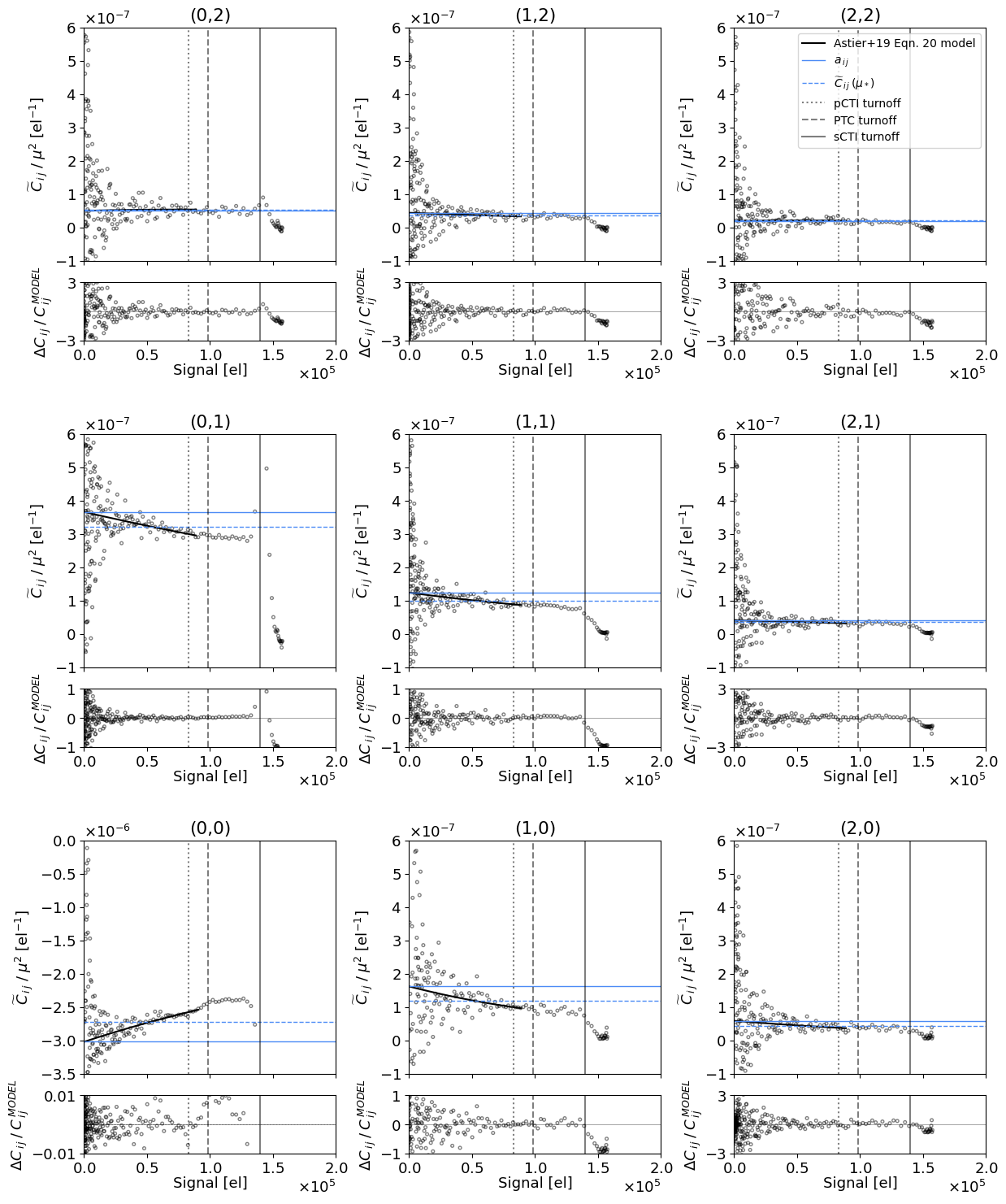}{0.95\linewidth}{(b) E2V Sensor (R03-S12)}}
    \caption{Continued.\label{fig:analytical_a_matrix_e2v}}
    \label{fig:analytical_a_matrix}
\end{figure*}

\subsection{Fitting the Covariance Function}\label{ssec:fittingthecovfunction}

We fit equation \ref{eq:cov_model} to our measured covariances from the zero signal limit to the pCTI turnoff for each channel in each of our sensors. We fit up to $O(\mathbf{a}^3)$ (the ``full covariance model") with an $8\times8$px covariance matrix ($i,j = 0, 1, \cdots7$) as a function of mean flat field signal.

Figure \ref{fig:charge_conservation_a_matrix} shows the zero-sum rule of $\mathbf{a}$ for different sizes of the matrix in sensor R03-S12 (ITL). We find that beyond 4\,px the fractional pixel area change fluctuations are largely dominated by noise, and that modeling out to 8 pixels, where $\sum a_{ij} < 1.04\times10^{-7}$ $\mathrm{el^{-1}}$, which is 5.8\% of $|a_{00}|$. Without including $\mathbf{b}$ in the fit, our sum rule is $\sum a_{ij} < 5.91\times10^{-9}$ $\mathrm{el^{-1}}$, which is only 0.345\% of $|a_{00}|$ and fully captures the zero-sum rule. The value of  $a_{00}$ varies by $3.42\%$ across amplifiers simply due to small differences in amplifier operation. Similar levels are reported for R24-S11 (E2V) in Table \ref{tbl:cov_fit}.

%In the later section \S\ref{ssec:isr}, we will discuss how we overcome this limitation by enforcing charge conservation. 

Figure \ref{fig:amatrix} shows the observed $a_{ij}$ matrix and profile of coefficients averaged over all channels., and the resulting fit parameters for both sensors are shown in Table \ref{tbl:cov_fit}. The $a_{ij}$ matrix shows a notable anisotropy between the cardinal directions of the pixel coordinate system ($a_{01} / a_{10} \sim 2.16$), as shown in Figure \ref{fig:amatrix}, indicating the BFE is stronger in the parallel direction. This anisotropy changes as a function of signal also due to the higher-order terms in equation \ref{eq:cov_model}. % This has the ultimate effect of causing the effective $a_{ij}$ to change as a function of flux.

The other parameter in these higher order terms, the $b_{ij}$ matrix, was previously measured to be sub-dominant by \citep{Astier_2019}, however we find that it is roughly the same order of magnitude as the $a_{ij}$ matrix, and is a sensitive variable in the full model fit.
%The $\mathbf{b}$ matrix models dynamic changes to the strength of the BFE, in other words changes to the effective impact of a single charge on a given pixel's effective area in the sub-pixel space during the integration of an exposure.
Figure \ref{fig:bmatrix} shows the fitted value of the $\mathrm{b}$ matrix for R03-S12 (ITL). We measure $b_{00} = -8.69\times10^{-7} \pm 1.11\times10^{-7}$, $b_{10} = -5.36\times10^{-6} \pm 2.24\times10^{-7}$, and $b_{01} = -4.71\times10^{-8} \pm 1.45\times10^{-7}$, all in units of $\mathrm{el^{-1}}$.  The element $b_{00}$ is $48\%$ of the magnitude of $a_{00}$. The $\mathrm{b}$ matrix has the effect of decreasing $a_{ij}$ if negative and increasing $a_{ij}$ if positive, corresponding to decreasing or increasing effective pixel areas, and it has the same units as the $a_{ij}$ matrix. It is not immediately clear how to physically interpret the value of the central $b_{ij}$ matrix elements or non-zero other elements of ${\bf b}$  other than as modifiers to the mathematical effect of $a_{ij}$ that enter in as higher-order terms of $\widetilde{C}_{ij}(\mu)$ at nonzero signal levels. That is to say $b_{ij}$ represents time-evolving (i.e., as charge accumulates) corrections to $a_{ij}$ in non-linear BFE components. 

Figure \ref{fig:analytical_a_matrix_itl} shows that for some $i, j$, $\widetilde{C}_{ij}(\mu) / \mu^2$ deviates from $a_{ij}$ for electron signal levels greater than zero, which must be due to contributions from the non-linear terms in equation \ref{eq:cov_model}. In correspondence with \citet{Astier_2019}, we find that terms higher order than $\mu^1$ are a significant component of the covariance function, accounting for as much as 15--30\% of the loss in pixel variance at $10^5$ $\mathrm{el}$, depending on the sensor tested, and are required for a good fit to the data. Excluding higher-order terms ($\mu^{4+}$) in the fit contributes a negligible error at least up to the pCTI turnoff level.  

The fit to our model leaves an average $\chi^2 / \mathrm{N_{dof}}$ of $13.7$ in each amplifier, which is larger $(\chi^2 / \mathrm{N_{dof}} = 14.9)$ when $\mathbf{b}=0$, which is unusually large as the model fit leaves $0.5\%$ fluctuation in the residual variance ($C_{00}$) above $5\times10^4$ ADU (bottom plots of Figure \ref{fig:analytical_a_matrix_itl}). The fluctuation has the same shape as a residual un-modeled non-linearity in the Analogue Signal Processing Integrated Circuit (ASPIC) \citep[described in][]{Astier_2019,Juramy2014}. The non-linearity takes on the form of a regular periodic function in signal level above $7.5\times10^4$ $\mathrm{el}$, and it might not have that much of an impact on our full covariance model if the surplus and deficit residuals balance the fit around the true PTC. However, by restricting the range of fluxes fit to those below the pCTI turnoff, we cut the fit range short enough that it might cause the non-linearity to bias the fit and give us a poor result. For this reason, the quality of the model fit is very sensitive to the flux range. We found that adjusting the maximum signal by $\pm10^3$ $\mathrm{el}$ (around the pCTI turnoff), we could achieve arbitrarily better $\chi^2$, but not necessarily a more accurate fit which models the BFE. On the other hand, the non-(0, 0) terms, which are also shown in Figure \ref{fig:analytical_a_matrix_itl}, exhibit good fits, with residuals smaller than the level of read noise and statistical fluctuations. By visual inspection of the PTC fit, it appears that the model accurately follows the PTC curve. In the next section, when we use the PTC model, we will discuss how we mitigate the impact of the poor NL correction.

Regardless, we find that the higher-order terms are non-negligible and result in non-linear alterations of $\widetilde{C}_{ij} / \mu^2$ as charge distributions grow over time.

\begin{figure}
    \includegraphics[width=0.48\textwidth]{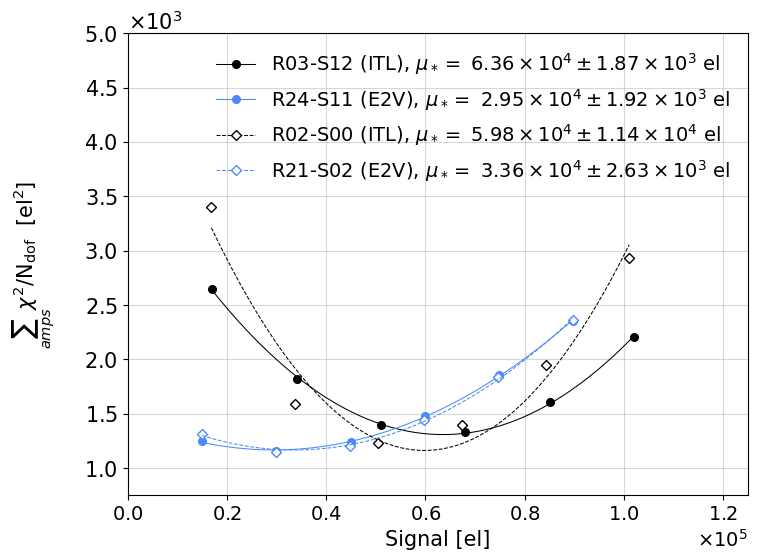}
    \caption{Residual $\chi^2$ values from a scan over the scale-conversion factor parameter space for all four sensors. We calculate the $\chi^2$ for each amplifier from the residuals below the pCTI turnoff between the corrected PTC and the expected linear behavior derived from the uncorrected PTC gain for each amplifier. We then sum all the $\chi^2$ values for each amplifier. A simple 3-parameter quadratic has been fit to the points across the parameter space to determine the factor that minimizes the $\chi^2$ statistic. This shows that we have a characteristic signal level that reconstructs the PTC, and this level is similar in sensors of the same type.}
    \label{fig:ptc_scan}
\end{figure}

\subsection{Correction of the BFE in Flat Fields}\label{ssec:correcting_ptcs}

\begin{figure*}
    \centering
    \gridline{
    \fig{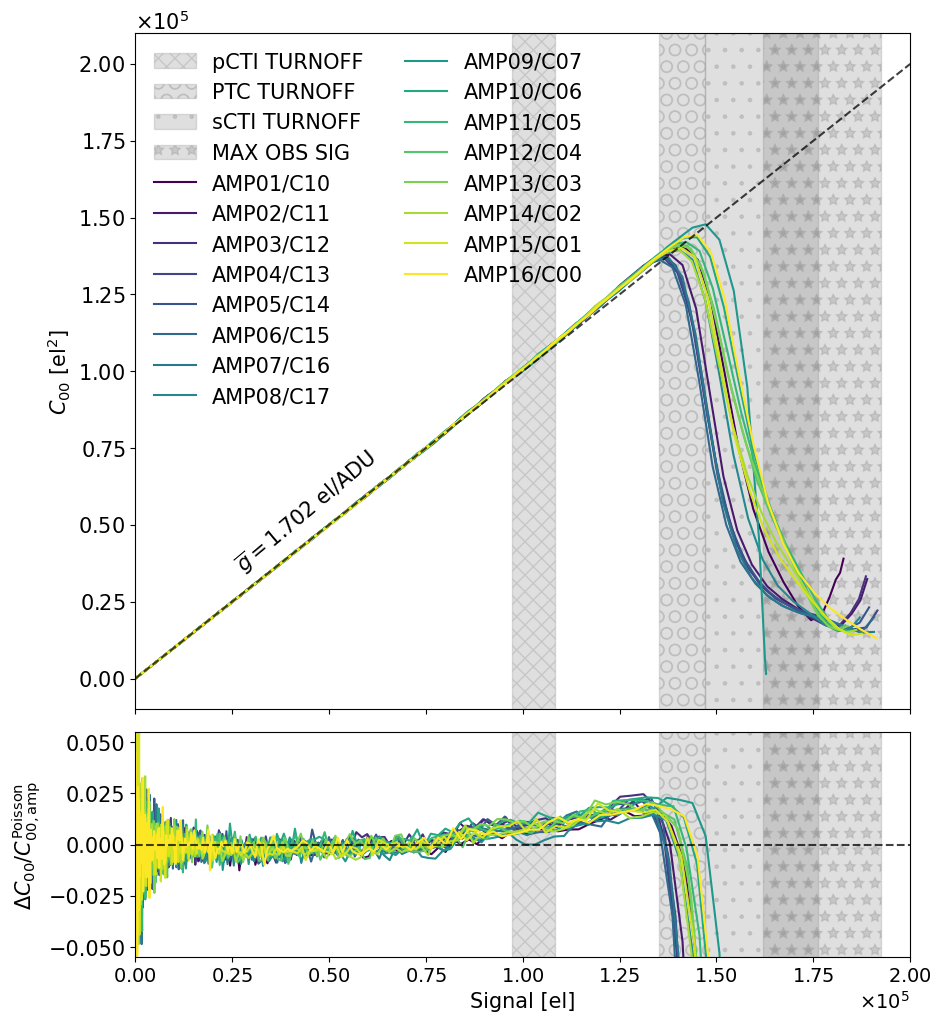}{0.49\linewidth}{(a) ITL Sensor (R03-S12)\label{fig:reconstructed_ptcs_itl}}
    \fig{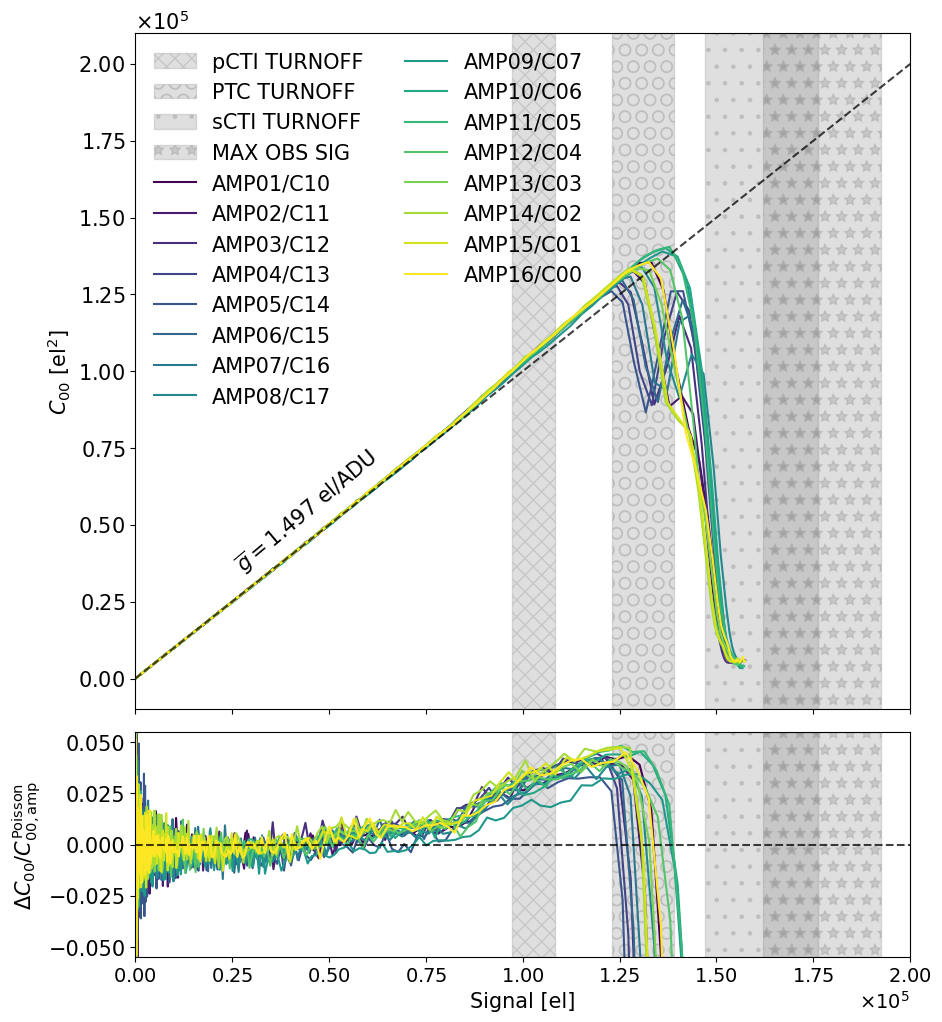}{0.49\linewidth}{(a) E2V Sensor (R24-S11)\label{fig:reconstructed_ptcs_e2v}}
    }
    \caption{Photon transfer curves reconstructed from corrected flat field images using the corresponding ideal kernels for the ITL sensor (R03-S12) on the left and the E2V sensor (R24-S11) on the right. On the bottom, we include the residuals from the expected linear behavior. Although we use a detector averaged kernel, we compute this for each of the 16 amplifiers. The average gain, included as a parameter in the post-correction full covariance model fit, increased by 0.35\% in the ITL sensor and 16\% in the E2V sensor.}
    \label{fig:reconstructed_ptcs}
\end{figure*}

In this section, we attempt to linearize the PTCs of our sensors by correcting the BFE out of the flat-field images using a kernel $K$ which includes higher-order BFE components. 

To include these higher order components into K, we construct the kernel (using equation \ref{eq:kernel}) by sampling $\widetilde{C}_{ij}(\mu)$ from our full covariance model at a non-zero signal level, which would therefore include higher-order effects. We will do this by: 
\begin{enumerate}
    \item Calculating $\widetilde{C}_{ij}(\mu)$ from the full covariance model at several signal levels.
    \item Use equation \ref{eq:kernel}, to calculate a kernel for each $\widetilde{C}_{ij}$.
    \item Correct the PTC using each kernel.
    \item Calculate the $\chi^2 / \mathrm{N_{dof}}$ between the corrected PTCs and the expected linear behavior defined by the gain of the uncorrected PTC.
    \item Determine the kernel that best corrects the PTC.
\end{enumerate}

\noindent This defines a characteristic signal level that can best correct the PTC (minimize the $\chi^2$). We will refer to this signal level as the ``ideal" signal level ($\mu_*$) and the corresponding kernel as the ``ideal" kernel.

We search a range of $\mu_* \in [\,10^4 \mathrm{ADU},\:\mathrm{pCTI\;turnoff}\,)$ in steps of $10^4$ ADU. This is the range within which we expect the $\chi^2$ curve as a function of signal level to follow a simple quadratic as it will not be significantly impacted by any systematic other than the BFE. We also only calculate the residual $\chi^2$-statistic from data below $5.0\times10^4$ $\mathrm{el}$ for E2V sensors and below $7.5\times10^4$ $\mathrm{el}$ for ITL sensors to avoid residual NL at higher signals from interfering with our test. We calculate the $\chi^2$-statistic from the variance component, exactly as \citet{Astier_2019}, which we described in \S\ref{ssec:ho_effects}.

We also take a few practical steps to ensure an accurate kernel. We normalized the kernel to enforce the zero sum rule (similar to charge conservation) by adjusting the central value of the central point (0, 0) of the kernel so that $\sum_{ij}{a_{ij}}=0$, for the measured correlations and the correlation model integrated out to infinity. We risk over-fitting noise if we calculate the kernel boundaries beyond 4 pixels, so we model the correlations beyond 4px away from the center of the kernel with an empirically measured power law, $C_{ij}/\mu^2 \propto r^{-\gamma}$ (typically $\gamma \sim 3.2$). However, we've observed that inclusion of the correlation model makes only a negligible difference to the kernel and the final correction. This central term ($K_{00}$) is typically decreased by approximately 10\% as a result of this zero-sum rule, and $K_{00}$ fluctuates across channels/amplifiers by approximately 10\% just due to differences in performance between amplifiers. Enforcing this zero sum rule has a large impact on the correction since the $(0, 0)$ pixel dominates the correction amplitude as we will show in the following sections. In order to avoid discontinuities at amplifier boundaries, we average the kernels per sensor and use the average in the final correction.

%\begin{figure*}
%    \centering
%    \includegraphics[width=0.9\textwidth]{figures/covariance_fit_residuals_R03_S12.png}
%    \caption{Residuals between the measured $2\times2$ pixel covariance matrix and the electrostatic model from \citet{Astier_2019}, equation 20, for the 16 channels in the ITL sensor (R03-S12). We show the mean of the residuals in 50 bins across the signal range. We fit up to the lowest PCTI turnoff of the channels.}
%    \label{fig:covariance_residuals_itl}
%\end{figure*}

%\begin{figure*}
%    \centering
%    \includegraphics[width=0.9\textwidth]{figures/covariance_fit_residuals_R24_S11.png}
%    \caption{Residuals between the measured $2\times2$ pixel covariance matrix and the electrostatic model from \citet{Astier_2019}, equation 20, for the 16 channels in the E2V sensor (R24-S11). We show the mean of the residuals in 50 bins across the signal range. We fit up to the lowest PCTI turnoff of the channels.}
%    \label{fig:covariance_residuals_e2v}
%\end{figure*}

Figure \ref{fig:ptc_scan} shows the result of a simple least-squares test which determines the best flux level to reconstruct the PTC. Most of the ideal signal levels are around half the range between zero signal and the pCTI turnoff, which could be a balance between underestimating the role of higher-order BFEs at low signal and potential deviations in the underlying assumptions of the \citet{Coulton_2018} correction scheme at higher signal levels. We also find that like-sensor types have similar ideal signal levels within fitting errors between our 4 sensors, and while the reason is not entirely clear, it could be the result of a physical parameter such as similar operating conditions or an artifact of the manufacturing \citep[e.g. similar doping levels in the substrate material as considered in][]{Rasmussen2016}.

Figure \ref{fig:reconstructed_ptcs} shows the ultimate result of applying the empirically determined best BF correction to the flat field images in order to reconstruct the linear PTC. The correction breaks down above the pCTI turnoff for each sensor. Since the kernels for each amplifier are averaged together, the amplifier with the lowest pCTI turnoff determines the maximum signal level that can be corrected. Above $7.5\times10^4$ $\mathrm{el}$ but below the pCTI turnoff, there is still some remaining uncorrected NL from the flat field projector, which prevents us from fully correcting the BFE above that signal level. This did not have an impact on our calculation of $\mu_*$ since we only used data points below this level in calculating our $\chi^2$ statistic specifically to avoid it interfering with our analysis.

We fit the full covariance model (up to the pCTI turnoff) to the BF-corrected PTC for both sensors. In the ITL sensor, the average $a_{00}$ decreased by $94.9\%$ and the residual anisotropy was corrected closer to unity ($a_{01}/a_{10} = 0.735$). For the E2V sensor, $a_{00}$ decreased by $97.1\%$ and the residual anisotropy was also corrected closer to unity ($a_{01}/a_{10} = 0.942$). This shows that the BFE correction is properly modeling the strength and anisotropy of the BFE in flat-fields below the pCTI turnoff.

\subsection{Correction of the BFE in Artificial Stars}\label{ssec:correcting_stars}

%\begin{figure}
%\includegraphics[width=\columnwidth]{figures/spot_shapess.png}
%\onecolumngrid
%\caption{This shows the distribution of spot shapes taken across all images in the Run 5 BF testing sequence. This metric of shape is both scale and rotation invariant. This shows both the uncorrected spot shapes as well as the shapes after the BF correction defined in \citep[Coulton_2018}.}
%\label{fig:spot_shapess}
%\end{figure}

%To identify the individual spots in these exposures, we used a technique that used set a simple signal threshold and identified clusters of pixels that were above this threshold. These regions were then classified as spots for further analysis, measurement, and classification. We find that this threshold method is sufficient to identify the majority of spots contained in the images without incorporating other sensor artifacts.

We also measure the BFE on artificial stars and apply the same ideal correction derived from flat fields. 
%The brighter fatter mechanism has the ultimate effect of flattening the contrast of images, and this consequence extends to images of stars (first observed by \citep{Lupton2014}). The BFE correlates pixels around stars, broadening their surface brightness with signal level. 
The BFE in stars could be a stronger effect than in the earlier case of flat fields since the contrasts (pixel-to-pixel differences) are larger and therefore the gradients of the charge distribution and the divergence of the deflection fields are also larger. 

\begin{figure}[h!]
    \includegraphics[width=\linewidth]{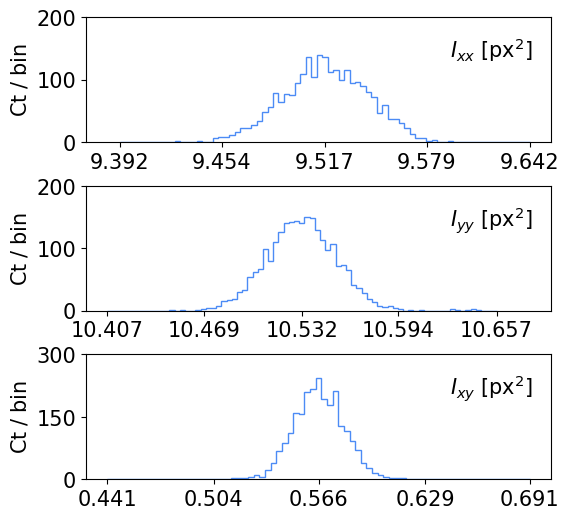}

    \caption{Distributions of the intrinsic shapes of the artificial stars used in these analyses (after selection cuts). Each distribution is centered on its mean and plotted with the same bin size. We derived these values from a collection of low-brightness images at the same exposure time (15\,s), and these distributions contain about 3200 spots. The distributions indicate a small average ellipticity ($a/b < 0.07$) and correlated orientation angle ($\bar{\theta} = -24.1$ deg) due to the relative alignment of the spot projector with the mask and focal plane, indicating that the major axes are almost parallel with the serial gate structure. Angles are given by $\tan(2\theta) = 2 I_{xy}/(I_{xx} - I_{yy})$ and measured relative to the serial transfer direction. }
    \label{fig:spot_shapes}
\end{figure}

To start, we found several problematic systematic effects in the images which needed to be addressed. Firstly, there is a wide illumination background in the artificial star images that peaks around $0.5\,e^-/\mathrm{px}/\mathrm{s}$, which is evident by eye in the top image of Figure \ref{fig:spotexample}. This is likely caused by excess transmission of light through the photo-lithographic mask. The background light level is much smaller than the Poisson noise of each star and is unlikely to significantly affect our photometry. However the shape-fitting algorithm is known to be sensitive to background light levels \citep{Hirata_2003, Refregier2012, Okura2018}, especially in the low signal-to-noise (SNR) regime, which will matter for our stars with peak signals below $2.5 \times 10^4\;\mathrm{e^-}$, so we model and subtract the background in each exposure using a 65px box size and 5px median filter, masking out the positions of the artificial stars. %More information about the process for background subtraction can be found in Appendix \ref{appendix:configs}.

Secondly, we observe Airy diffraction rings on each of the stars due to the pinholes of the spot projector mask, which can be seen in the bottom image of Figure \ref{fig:spotexample}. However, as we will discuss further, we will study the differences between spot sizes at any signal level with respect to the same star's intrinsic shape at low signal and the impact of the detailed structure will be cancelled out even if the diffraction rings of different spots have non-negligible overlap. 

Thirdly, The focal-plane-to-projector orientation and optical aberration from the lens caused minor distortions of the artificial stars, especially away from the center of the grid, and they cannot be perfectly parameterized by isotropic 2D Gaussians. We observed that the central and brightest stars of the grid hardly vary in intrinsic shape between exposures at the same or different projector positions. To avoid any other systematic effect related to the intrinsic alignment of the spots themselves, we therefore select only the top 5\% brightest stars in total flux in each image. These stars happen to correspond to the same pinholes in the spot projector between in each image. In order to avoid including stars that land on the edges of the sensor, which skews their shapes, we also clipped stars beyond $3\sigma$ of the mean at each exposure time. This cut only removed O(10) data points in a few of the exposures. We ultimately selected approximately 80 spots in each exposure for our analysis. With 40 images at each exposure level, we use approximately 3200 spots/exposure level. 

Figure~\ref{fig:spot_shapes} shows the approximate intrinsic shapes of the artificial stars used in this analysis, which we derive from their shapes at 15s of exposure, which corresponds to a relatively low peak signal level of $2.5\times10^4 \mathrm{e^-}$. The distortions are correlated across the grid with the major axes are consistently 7\% larger than the minor axes, and the major axes of all the spots are aligned at a $-24\;\mathrm{deg}$ angle relative to the serial readout direction (horizontal and to the left in Figure \ref{fig:spot_shapes}). We do not believe this affects our result, as we will discuss further in \S\ref{ssec:nonzerocurl}.  

%To investigate the size growth caused by the BFE in this scenario, we only need to look at the growth of the second moments of shapes with signal level on our sample of artificial stars.
In Figure \ref{bf_spots_itl}, we show the growth of all three independent components of the second-moments matrix ($I_{xx}$, $I_{yy}$, $I_{xy}$) separately for each sensor type. We chose the mean star shapes derived from the 15-second exposures as a fiducial point from which to show the growth of the size that result from the BFE since the lowest flux stars suffers from low signal-to-noise ratio due to the extended background. We therefore measure the size-growth of each star, corresponding to an individual hole in the projector mask, relative to its intrinsic shape. We also use the peak signal of each spot instead of the integrated signal as a measure of the signal level so that we can relate our spot photometry to different measurements of the pixel full-well capacity.

\begin{figure*}[hp]
    \centering
    \gridline{\fig{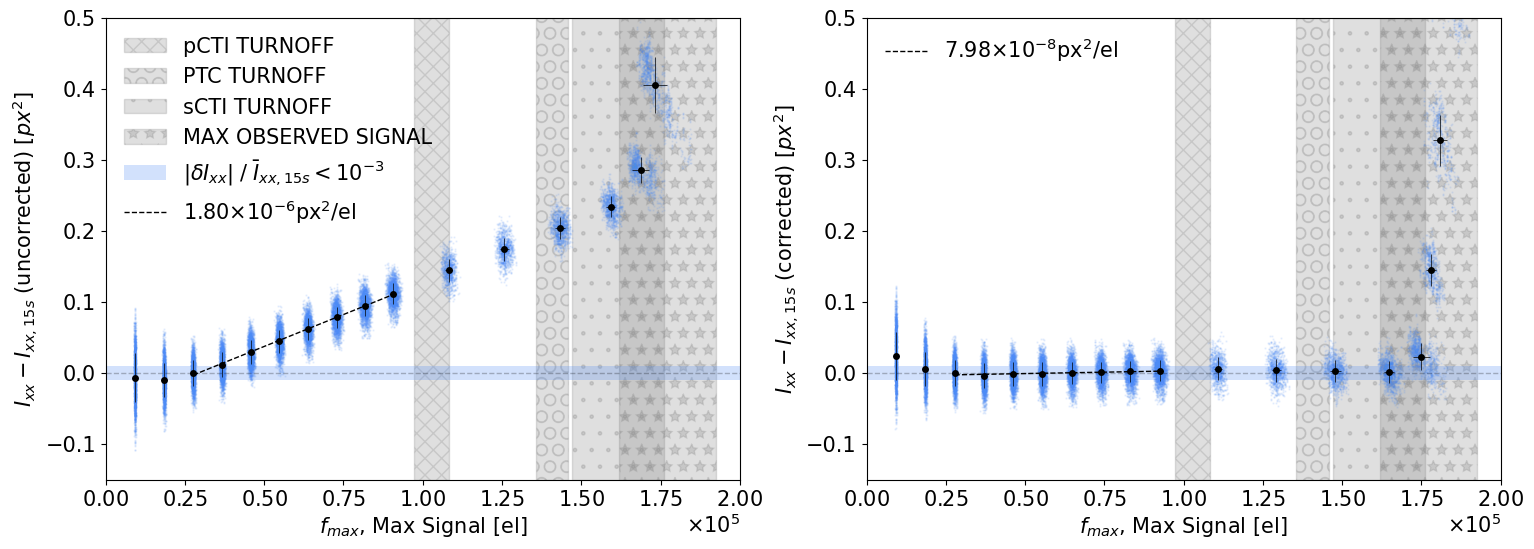}{.95\textwidth}{}}
    \gridline{\fig{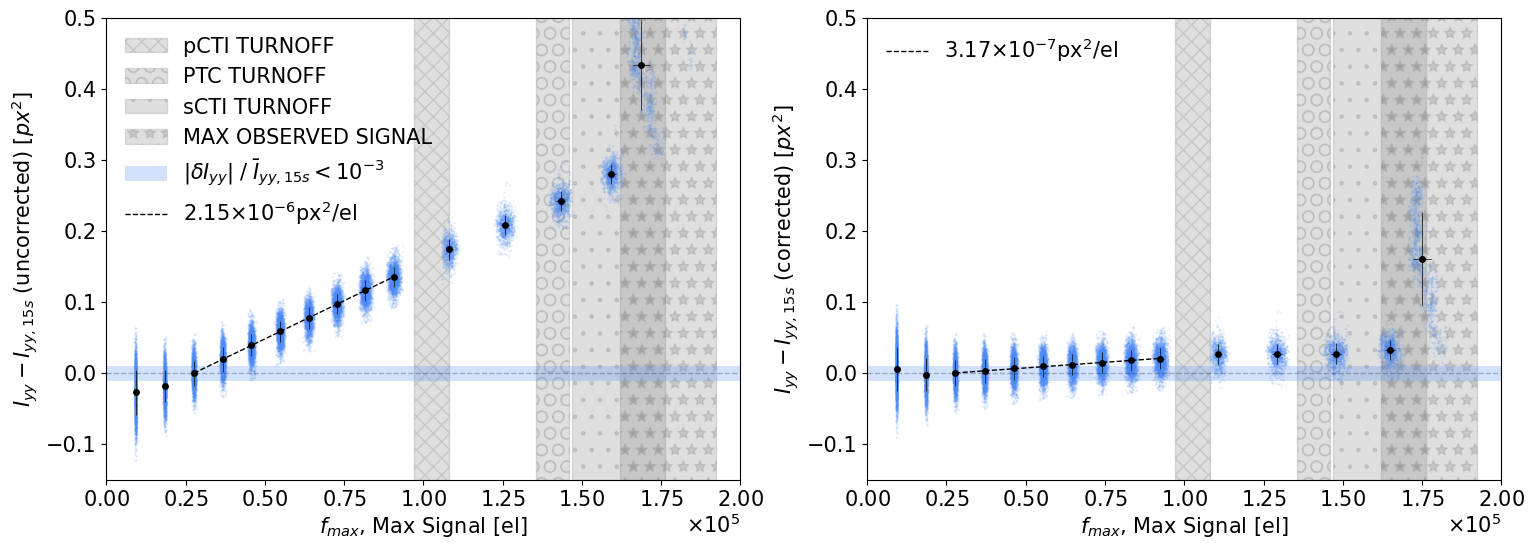}{.95\textwidth}{}}
    \gridline{\fig{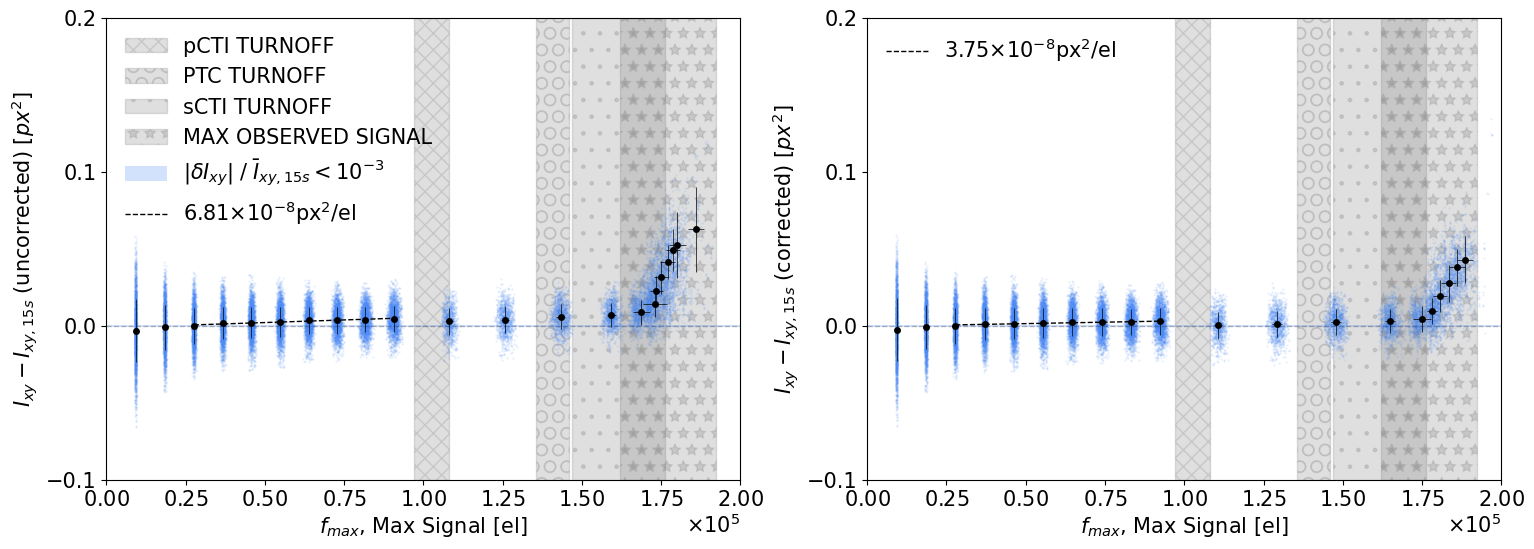}{.95\textwidth}{(a) ITL (R03-S12) \label{bf_spots_itl}}
    }
    \caption{The second moments of the artificial stars taken from images vs. their peak pixel value for an ITL sensor (R03-S12). Left column is the uncorrected data and the right column is corrected data. The rows correspond to $I_{xx}$, $I_{yy}$, and $I_{xy}$ from top to bottom. Each blue point is a single spot from a single exposure. Error bars are standard errors of the distribution about the mean at each exposure time in moment and flux, and a line has been fitted to the 15--50\,s points, which corresponds to $2.5-9.0\times10^4$ $\mathrm{el}$ (just below the pCTI turnoff). The vertical shaded regions represent the range of full-well capacities of the 16 amplifiers on the sensor using the four different methods. The blue shaded region is the boundary to model the shape to within 1 part-per-thousand.}\label{fig:spots}
\end{figure*}
\begin{figure*}[hp!]\ContinuedFloat
    \medskip
    \centering
    \gridline{\fig{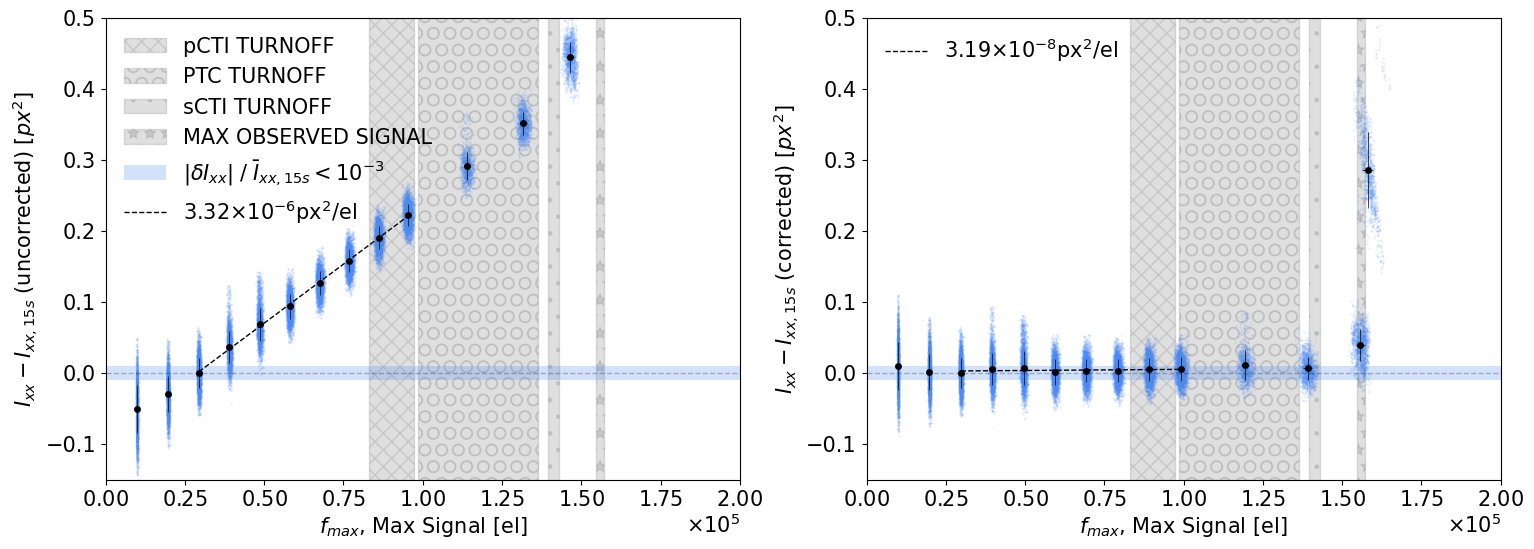}{.95\textwidth}{}}
    \gridline{\fig{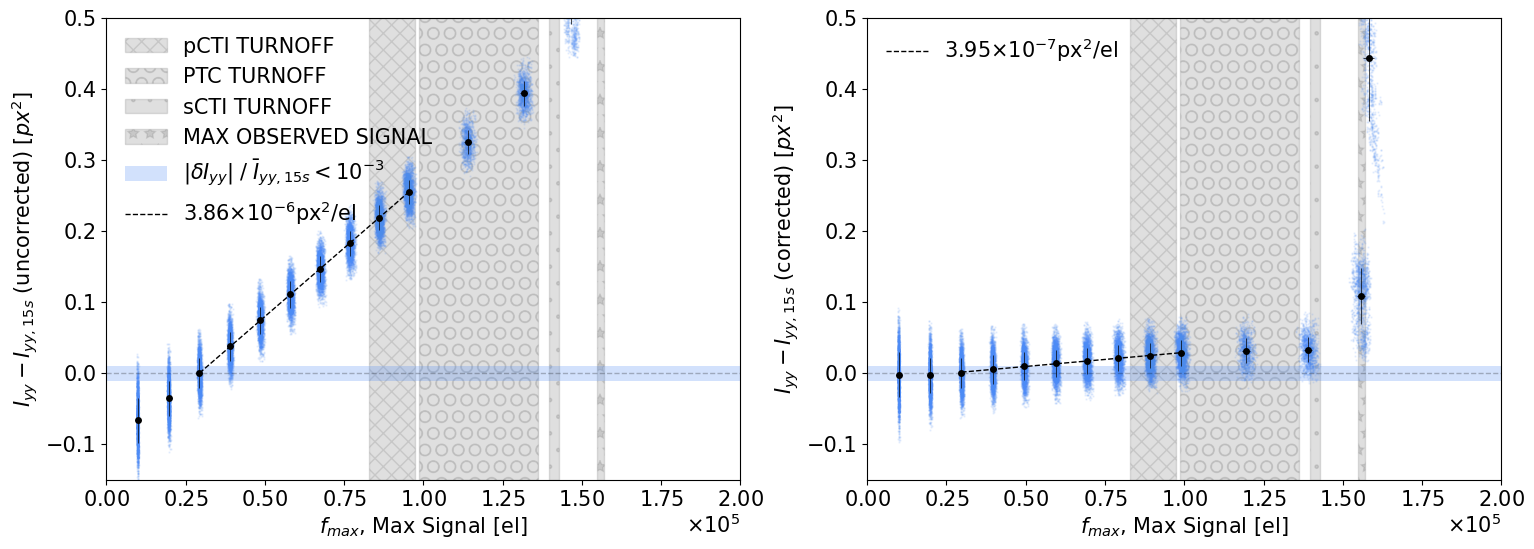}{.95\textwidth}{}}
    \gridline{\fig{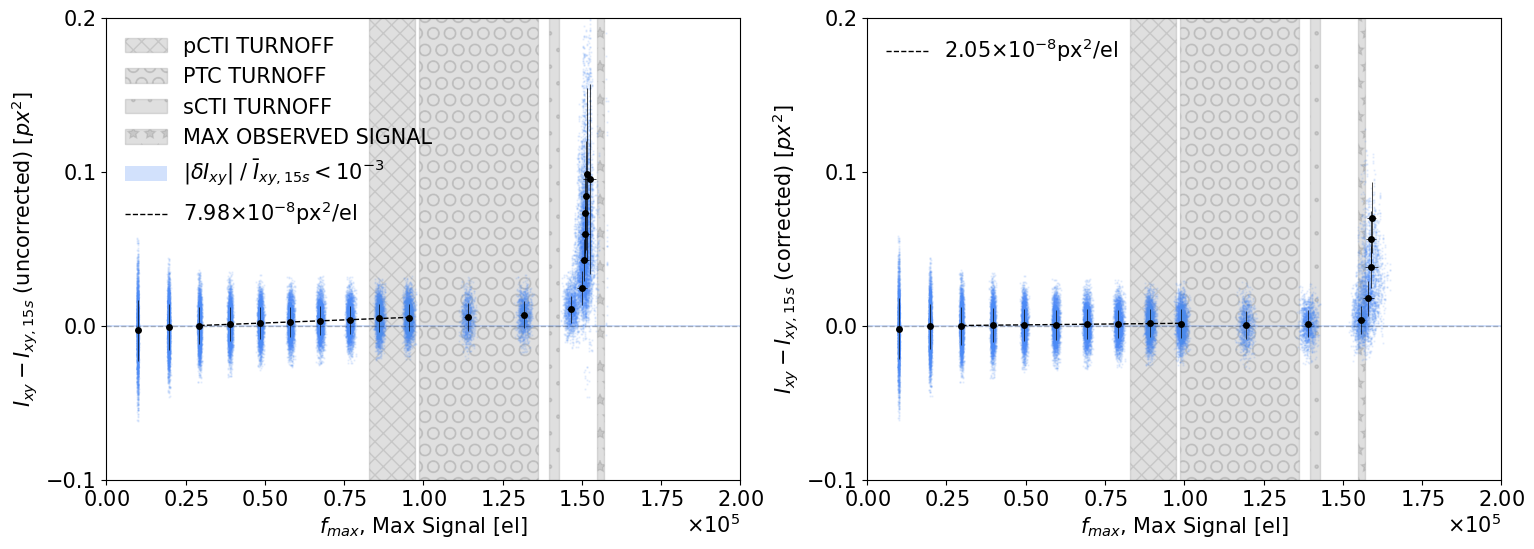}{.95\textwidth}{(b) E2V (R24-S11) \label{bf_spots_e2v}}}
    \caption{Continued.}\label{bf_spots}
\end{figure*}

\begin{table*}
    \label{tab:spot_tbl}
    \centering
    Summary of BF Correction on Artificial Stars
    \begin{tabular*}{\linewidth}{@{\extracolsep{\fill}}lcccccc}
      \toprule[1pt] % <-- Toprule here
      \midrule[0.3pt]
      Sensor & $\alpha_{xx}$ & $\sigma_{xx}$ & $\alpha_{yy}$ & $\sigma_{yy}$ & $\alpha_{xy}$ & $\sigma_{xy}$\\
      & $[\mathrm{px}^2 / $el$]$ & $[\mathrm{px}^2 / $el$]$ & $[\mathrm{px}^2 / $el$]$ & $[\mathrm{px}^2 / $el$]$  & $[\mathrm{px}^2 / $el$]$  & $[\mathrm{px}^2 / $el$]$ \\
      \midrule[0.3pt] % <-- Midrule here
      R03-S12 (ITL) & $ 1.80 \times 10^{-6} $ & $ 2.65 \times 10^{-8} $ & $ 2.15 \times 10^{-6} $ & $ 5.34 \times 10^{-9} $ & $ 6.81 \times 10^{-8} $ & $ 7.59 \times 10^{-9} $  \\ 
                    & $ 7.99 \times 10^{-8} $ & $ 2.32 \times 10^{-8} $ & $ 3.17 \times 10^{-7} $ & $ 7.66 \times 10^{-9} $ & $ 3.75 \times 10^{-8} $ & $ 7.39 \times 10^{-9} $ \\
                    & (95.6\%) & $-$ & (85.2\%) & $-$ & (44.9\%) & $-$ \\
      \midrule[0.3pt]
      R24-S11 (E2V) & $ 3.32 \times 10^{-6} $ & $ 4.10 \times 10^{-8} $ & $ 3.86 \times 10^{-6} $ & $ 6.03 \times 10^{-9} $ & $ 7.98 \times 10^{-8} $ & $ 2.53 \times 10^{-9} $ \\ 
                    & $ 3.19 \times 10^{-8} $ & $ 3.76 \times 10^{-8} $ & $ 3.95 \times 10^{-7} $ & $ 1.29 \times 10^{-8} $ & $ 2.05 \times 10^{-8} $ & $ 2.42 \times 10^{-9} $ \\
                    & (99.0\%) & $-$ & (89.8\%) & $-$ & (74.3\%) & $-$ \\
      %\midrule[0.3pt]
      %R02-S00 (ITL) & 0.0117 & 0.0133 & 0.0006 \\ 
      %&  -0.0033 & -0.0026 & 0.0004 \\
      %& (71.79\%) & (80.40\%) & (33.33\%) \\
      %\midrule[0.3pt]
      %R21-S02 (E2V) & 0.0248 & 0.0266 & 0.0006 \\ 
      %&  -0.0017 & -0.0015 & 0.0002 \\
      %& (93.15\%) & (94.36\%) & (66.67\%) \\
      \bottomrule[0.3pt] % <-- Bottomrule here
    \end{tabular*}
    \caption{Strength of the BFE before (top) and after (middle) correction, with the relative correction level (bottom) on artificial stars in different sensor types. Strengths are given as the fitted slope of the line from equation \ref{eq:lin_fit} with $1$-$\sigma$ fitting errors. A value $\alpha_{\mu\nu} > 0$ means the correction is undercorrecting the BFE.}\label{spot_table}
\end{table*}

To quantify the magnitude of the BFE, we determine the growth of the second moment with respect to this peak signal level. We fit the data points above the influence of our background light ($f_{\mathrm{max}} > 2.5\times 10^4$) and below the pCTI turnoff (which is our linear size-growth regime) with the function:
\begin{equation}\label{eq:lin_fit}
    I_{\mu\nu} - I_{\mu\nu, 15\mathrm{s}} = \alpha_{\mu\nu}f_{max} + c_{\mu\nu},
\end{equation}
where the indices run over the $\mathbf{x}$ and $\mathbf{y}$, and $\alpha_{\mu\nu}$ determines the relative strength of the BFE, and $c_{\mu\nu}$ just captures the difference between our fiducial point at $I_{\mu\nu, 15\mathrm{s}}$ and the true intrinsic shape of our spots in the limit of zero signal, which is important for fitting but unimportant for deriving the strength of the BFE.

For the ITL sensor, the size of our artificial stars grow by 1.75\% from their intrinsic size ($T_{PSF} = I_{xx} + I_{yy}$) near the pCTI turnoff, and they grow by 3.25\% for the E2V sensor. However, this size growth is asymmetric, just as we observed in the case of flat fields. For the sensors we measured, the BFE ($\alpha_{\mu\nu}$) is consistently 16--19\% stronger in the parallel direction than in the serial direction for these stars.

% Describe the correction
Figure \ref{bf_spots_itl} also show the result of applying the correction derived using the ideal kernel. We correct $T=(I_{xx} + I_{yy})$ by 89.9\% in ITL and 94.1\% in E2V, which is worse than the correction on $a_{00}$ in flat fields (95-97\%), and the correction preserves 67.2-67.8\% of the anisotropy in the BFE.  The fact that the correction of $I_{xy}$ is worse than the other components is likely an artifact of it being a small in magnitude. Table \ref{spot_table} breaks down the detailed fit parameters between the two sensors we studied.

%level is also asymmetric in the parallel and serial components. 

We also show the different turnoff points of pixel full-well discussed in \S\ref{ssec:fullwellcapacitysection} to show that the BFE has impacts over the full dynamic ranges of our sensors. Interestingly, we observe that the PTC turnoff has little effect on the shapes of the corrected spots in our study compared to the physical effects at the sCTI turnoff and the pCTI turnoff, which results in a small deviation from the linear relation as shape of the spots increases due to deferred charge above this level.

%\begin{figure*}
%  \newpage
%  \begin{subfigure}[t]{.49\textwidth}
%    \centering
%    \includegraphics[width=\linewidth]{figures/T-ellipse-680nm-R03_S12.png}
%  \end{subfigure}
%  \hfill
%  \begin{subfigure}[t]{.49\textwidth}
%    \centering
%    \includegraphics[width=\linewidth]{figures/T-ellipse-680nm-R24_S11.png}
%  \end{subfigure}
%
%\caption{The mean of the trace 2nd order PSF moment matrix ($T = I_{xx} + I_{yy}$) of centrally projected elliptical stars (normalized to a 15s exposure spot) are shown relative to their mean peak pixel value for both an ITL sensor (R03-S12) and an E2V (R24-S11) sensor. Each point is a single spot from a single exposure. Error bars are errors of the mean in moment and flux and are shown for each exposure population (represented by the color bar), and a line has been fitted to the 15-50s points. The vertical shaded regions represent the range of full-well capacity of the 16 amplifiers on the sensor using four different methods.}
%\label{fig:bf_T_ellipse}
%\end{figure*}

\subsection{Charge Conservation}\label{ssec:chargecons}

\begin{figure}
\includegraphics[width=\linewidth]{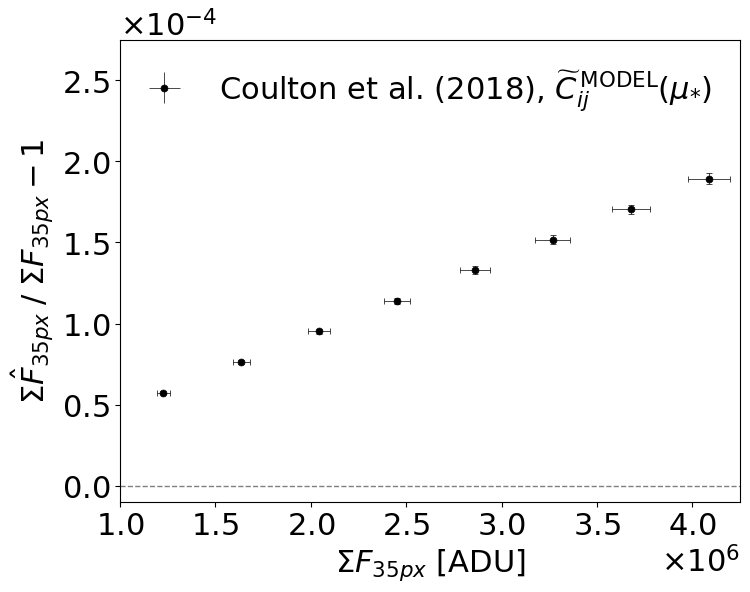}
\caption{The quantity $f = \sum_{r < 35\mathrm{px}}{\hat{F}} / \sum_{r < 35\mathrm{px}}{F}$ is the ratio of integrated flux within 35px of the centroid of each spot, after/before the BFE correction is applied on each of the  spots in the ITL sensor (R03-S12) data. We only plot spots with peak signals below the pCTI turnoff. Error bars are standard errors of the distribution at each exposure time.}
\label{fig:chargeconservation}
\end{figure}

The charges are not lost when they are deflected, and a good physically motivated correction should conserve charge. While we implement the zero sum rule on $a_{ij}$ and a zero-sum condition on the whole of $K$ by adjusting the (0,0) term \citep[the way][defines the algorithm, and as it was previously implemented in the LSST science pipelines]{Coulton_2018}, the correction only conserves charge in the continuous limit ($\left \langle \delta F \right \rangle = 0$). It does not conserve charge on small scales due to the application of the kernel in equation \ref{bfcorrectioneqn}
%As discussed in \S\ref{ssec:isr}, the original implementation of equation \ref{bfcorrectioneqn} did not conserve charge. 
Figure \ref{fig:chargeconservation} shows the difference in the measured charge of stars after applying the correction. We used 35px pixel aperture for photometry, which is a large enough aperture compared to the PSF size (6px FWHM) to capture the total footprint of each star without needing to assume a particular shape for the profile. The brightest unsaturated spots (below the pCTI turnoff) have a small surplus $\delta < 0.02\%$ of total integrated charge compared to the same star in the uncorrected image. 
%The new correction by fully conserves charge by construction, which is obviously a more physically motivated model. 
Charge non-conservation was more apparent in high-contrast images such as our artificial stars than in flat-fields, which would suggest that the error is the result of a local deviation of the correction from Gauss's Law as high-contrast images have larger deflection field divergences in the application of the correction. Improvements which enforce Gauss's law locally (on each pixel boundary) are are currently in development and undergoing a detailed study.

%Figure \ref{fig:concl_spots} shows the results of using the previous implementation of the correction (labeled ``$\widetilde{C}_{ij}^{\mathrm{MODEL}}(\mu_*)$, Coulton+2018") and the new ``flux conserving" correction (labeled ``$\widetilde{C}_{ij}^{\mathrm{MODEL}}(\mu_*)$") on the artificial stars for sensor R03-S12 (ITL), indicating that the correction level as measured by the residual strength of the BFE also improves, likely due to better local modeling of the displacement fields.

%While we implement the zero sum rule on $a_{ij}$ and a zero-sum condition on the whole of K by adjusting the (0,0) term  \citep[the way][defines the algorithm, and as it was previously implemented in the LSST science pipelines]{Coulton_2018}, the correction does not conserve charge on small scales due to the application of the kernel in equation \ref{bfcorrectioneqn}, and an improvement was made to the application of the kernel via equation \ref{bfcorrectioneqn} by enforcing Gauss's law locally (on each pixel boundary) to avoid this resulting in a loss of charge conservation across the image. This has been added into the LSST Science Pipelines as the ``\texttt{fluxConservingBrighterFatterCorrection()}'' in \texttt{ip-isr/isrFunctions.py}, which is in the codebase referenced by \citet{LsstDMPipeline1}. A detailed study of the performance of each implementation will be discussed in \S\ref{ssec:chargecons}.

\section{Discussion}\label{sec:discussion}
\subsection{Testing the Assumptions of Scalar BFE Theory}
In this section, we examine the assumptions that allow $\widetilde{C}_{ij}(\mu)$ to take the form $\mu^2 \nabla^{2} K$. The \citet{Coulton_2018} approach assumes (1) that there are no higher-order terms that result in non-linear BFE, and (2) the deflection field produced by accumulated charge in a pixel is conservative (zero curl).

\subsubsection{Non-Linear BFE Components}

\citet{Coulton_2018} defines the kernel (equation \ref{eq:kernel}) such that the gradient along one direction ($\nabla_{i} K$) represents the transverse displacement field along that direction. The impact of the BFE on flat-field covariances is the expected Poissonian behavior modified by the change in pixel area so that $\widetilde{C}_{ij}\;/\;\mu^2$ models an \textit{effective} fractional area displacement matrix, and in the limit of zero signal $\widetilde{C}_{ij}\;/\;\mu^2 \rightarrow a_{ij}$ \citep[the full covariance model parameter of ][]{Astier_2019}. 
%The gradient of the kernel along one boundary is similar to a displacement field along that boundary, and 

The implication of a fixed kernel $K$ in the \citet{Coulton_2018} model is that $\widetilde{C}_{ij}\;/\;\mu^2$ does not change with signal level. Figure \ref{fig:analytical_a_matrix} shows the limited validity of the assumption that $\widetilde{C}_{ij}\;/\;\mu^2$ is independent of signal level. We show in the figure that $\widetilde{C}_{ij}\;/\;\mu^2$ agrees with the $a_{ij}$ matrix by \citet{Astier_2019} in equation \ref{eq:cov_model} only in the limit of zero signal, and that near the pCTI turnoff it deviates by 20--30\%, depending on the sensor. 
%The deviation from the assumption of $\widetilde{C}_{ij}\;/\;\mu^2 = \mathrm{const.}$ is the natural consequence of the \citet{Astier_2019}'s model that assumes the accumulated charge flow between neighbouring pixels.

%In order to avoid misattributing the impact of BF to impacts from other sources of pixel-correlations, we limit the range of our full covariance model fit to low signal levels. From our flat-field calibrations, we construct our PTC and fit our covariance model to measure the size of the BFE and construct the correction kernel.  
%

The deviation in the PTC is well-modeled by the higher order (non-linear) terms of equation \ref{eq:cov_model}. The main contributing higher-order term of the full covariance model has two components ($\mathbf{a}\otimes\mathbf{a} + \mathbf{ab}$) with two parameters ($\mathbf{a}$ and $\mathbf{b}$). The $\mathbf{a}$ component is the principal BFE component and the $\mathbf{b}$ component strengthens (positive element) or weakens (negative element) the pixel boundary shifts with growing charge accumulation. The $\mathbf{b}$ parameter is mostly negligible at low flux as the fit quality our our calibration data to equation \ref{eq:cov_model} is slightly worse if we fix b to zero than if we leave it as a free parameter (per-amplifier average reduced weighted $\chi^2 / \mathrm{N}_{\mathrm{dof}} = 13.7$ with $\mathbf{b}$ and $\chi^2 / \mathrm{N}_{\mathrm{dof}} = 14.9$ without \textbf{b}). While both fits had some residual NL, the primary difference between the fits occurs at high signal levels (near pCTI turnoff). However, \citet{Coulton_2018} do not distinguish between \textbf{a} and \textbf{b} or include their higher order components.

The physical meaning of $\mathbf{b}$ is not as clear as $\mathbf{a}$. The negative $b_{00}$ element could have the same impact as a ``space-charge effect" as the drift field decreases due to the accumulated charge and therefore increases the drift time of the electron and the time that the BFE has to act on those incoming charges as they drift into the pixel. The effective area of the central pixel effectively decreases due to the space-charge effect and would therefore produce negative best-fit values of $b_{00}$.

The non-(0,0) components of $\mathbf{b}$ are also negative except for the parallel terms, which could also be the result of the space-charge effect interfered with by residual pCTI. These elements are also asymmetric, with the best-fit measurement of $b_{10}/b_{01} = 2.82$, which could result from a growing asymmetry in pixel boundary shifts with charge accumulation. As charge accumulates, if the stored charge cloud distorts anisotropically, further charges would preferentially be deflected in a particular direction as a type of charge feedback effect.

Figure \ref{fig:concl_spots} summarizes the correction of artificial stars in R03-S12 (ITL) for different kernels evaluated from different signal levels of the full covariance model, which encode the influence of varying amounts of the higher-order BFEs. The correction based on using $\widetilde{C}_{ij}(0) \sim a_{ij}$ over-corrects the artificial star images, and the correction based on $\tilde{C}_{ij}(\mu_{\mathrm{pCTI}})$ under-corrects the images. Some kind of mechanism to alter the BFE from the pixel boundary shift at non-zero signal level is needed. An adjustment of the strength is needed by constructing a kernel from the measured covariance model at some higher signal level. Scanning the whole covariance model to find the ideal signal level that reconstructs the expected flat-field behavior (as we showed in \S\ref{ssec:correcting_ptcs}) is the best method to balance the assumptions of the kernel approach and the higher-order BF contributions. 

The ideal kernel however, still does not fully correct the BFE in flat fields or artificial stars (Figures \ref{fig:reconstructed_ptcs} and \ref{fig:spots}). The correction also leaves differing residuals in different sensors (Table \ref{spot_table}). It could be some error in enforcing the zero-sum rule, some other unmodeled source of measured pixel correlations (such as NL), or we could be missing some component in the physical model of the BFE in the correction. \citet{Astier_2019} pointed out that all of the correction methods still leave on the order of 10\% of the initial effect in the images \citep{Guyonnet2015,gruen2015} and that all of the proposed correction techniques assume no contribution from higher-order terms.

\subsubsection{Non-Zero Curl Components}\label{ssec:nonzerocurl}

\begin{figure*}
    \centering
    \includegraphics[width=0.9\textwidth]{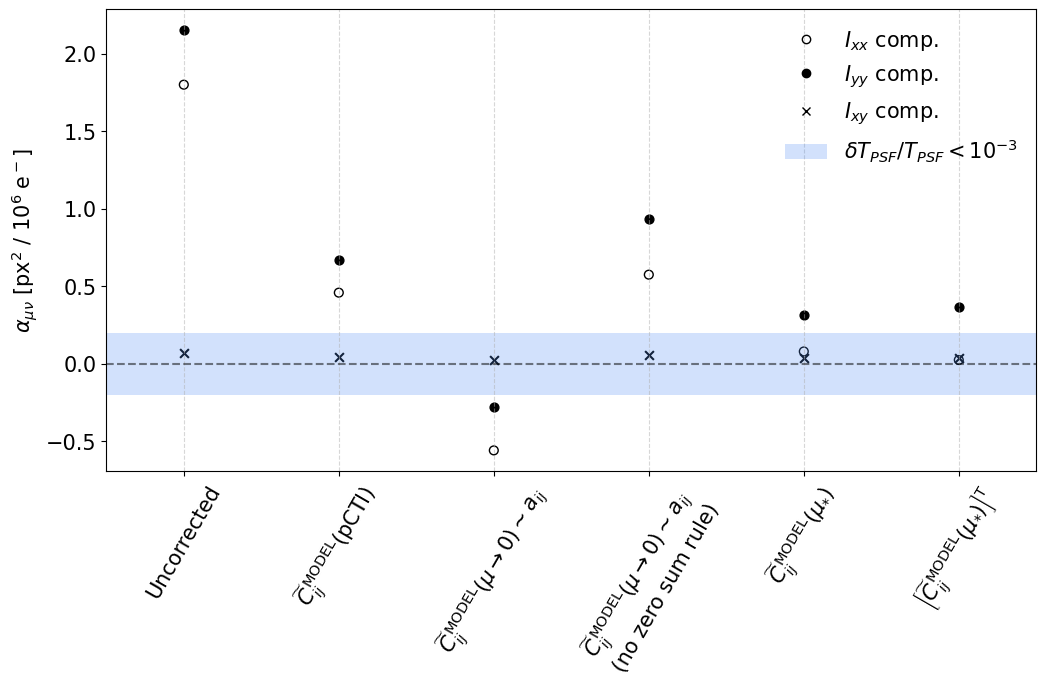}
    \caption{The strength of the BFE measured from artificial stars (fit to equation \ref{eq:lin_fit}) before/after applying scalar corrections derived from different configurations of $\widetilde{C}_{ij}$ for sensor R03-S12 (ITL). We show the fitted slopes for the uncorrected artificial stars as well as for the artificial stars corrected using the covariance model sampled at zero signal, sampled at the pCTI turnoff (minimum value of all amplifiers), the derived ideal signal level from our full PTC scan, and for the last case, we also show the correction with the kernel transposed (purposefully applied in the wrong orientation), which shows how sensitive the correction is to the fixed-plane BF anisotropy. The shaded area represents the total allowed residual BFE ($\alpha_{xx} + \alpha_{yy}$) to meet the LSST science requirements \citep{lsst_SciDoc} on $T_{PSF} = I_{xx} + I_{yy}$ to the pCTI saturation limit ($10^5\,\mathrm{el}$) for stars of size $\left \langle T_{PSF} \right \rangle = 20$ px$^2$, under the assumption that our artificially produced stars are real, measured PSF stars.}
    \label{fig:concl_spots}
\end{figure*}
% For the last case, we show the result of applying the original implementation of the \citet{Coulton_2018} algorithm (``Coulton+18"), the flux conserving correction described in \S\ref{ssec:isr}, and the flux conserving
The correction algorithm in \citet{Coulton_2018} is entirely based on scalar products and therefore no vector-like curl-component of the deflection field is included or correctable in this model. The correction in Equation \ref{bfcorrectioneqn} only contains a divergence component, however on small scales there could be a non-negligible curl-component. A non-zero curl component could be generated from the intrinsic anisotropy that exists in the sensors. 

The validity of the zero-curl assumption is directly tested by comparing the correction of anisotropy in flat field covariances and artificial stars. Flat field pixel correlations have a strong dependence on the divergence component of the BFE and a comparatively small dependence on the curl component if any exists due to the spatial uniformity of the charge distribution. Pixel correlations in artificial star images by contrast could depend strongly on a curl component of the BFE. The BFE in the parallel direction is consistently 215-234\% larger than it is in the serial direction when measured from the $a_{01}/a_{10}$ in flat fields.  And the parallel direction is consistently 16-19\% stronger when measured by the parallel second moment of the artificial stars.

In (Figure \ref{fig:concl_spots}), only about 32\% of the anisotropy in stars gets corrected by the algorithm. On the other hand, the anisotropy in flat field covariances is corrected by 60-65\%. The \citet{Coulton_2018} algorithm corrects anisotropy better in the case of flat fields simply because it better models the impact of local physics on flat field statistics, where the curl component is smaller.  

This could also impact the total correction level between flat fields and stars. In the application of the kernel to artificial stars, the correction reaches the total convergence condition before it can properly redistribute charge in the most adjacent pixels that capture the anisotropy, whereas in flat fields, with small pixel contrasts, the correction in the adjacent pixels makes up a larger total fraction of the charge redistribution at each iteration. The correction in flat fields therefore redistributes more charge in the same number of iterations before meeting the convergence condition than it does in artificial star images. If we take an extreme case and purposefully apply the transposed kernel to the raw images--that means applying the correction with the anisotropic component of the kernel in the wrong direction (Figure \ref{fig:concl_spots})--the final anisotropy is roughly unchanged with the uncorrected case. This suggests that in artificial star images, most of the correction is dominated by the central pixels of the kernel, however we can see from charge conservation in the $\mathbf{a}$ matrix (Figure \ref{fig:charge_conservation_a_matrix}) that most of the charge redistribution that occurs is due to the pixels that are not immediately adjacent to the central pixel. This is further evidence that the curl component is not negligible. 

An improvement to the algorithm might then consider a vector approach, which deals with each pixel boundary. However, there are not enough degrees of freedom to derive all four boundary shifts on a pixel from flat field statistics alone, even accounting for shared boundaries ($a_{i,j}^N = a_{i,j+1}^S$ and $a_{i,j}^E = a_{i+1,j}^W$), so this method would require further modeling with each boundary left as a free parameter.  Work is ongoing in the development and testing of other approaches like this \citep[such as the one proposed by][]{Astier_2023}.

\begin{figure}
    \includegraphics[width=\linewidth]{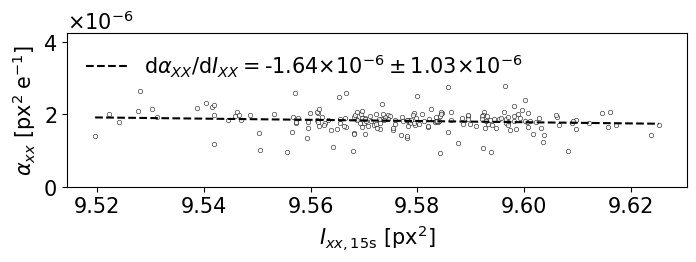}
    \includegraphics[width=\linewidth]{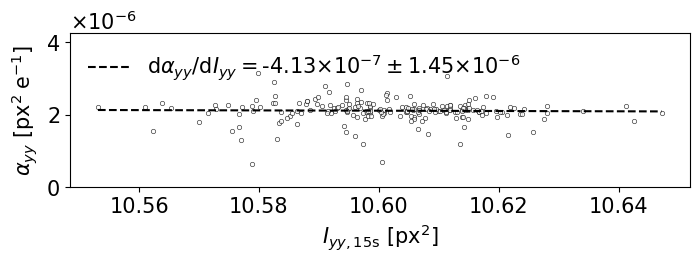}
    \includegraphics[width=\linewidth]{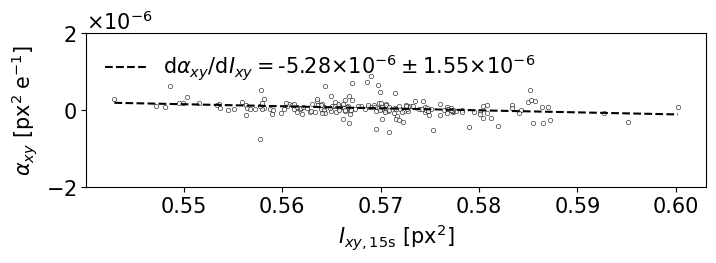}
    \caption{The decrease in the influence of the BFE in more extended sources, which would have smaller flux gradients in any direction of the pixel plane. We show the slope of the BFE by component, derived from the 81 individual spots used in our analysis, where each point corresponds to a single hole in the spot mask, relative to their intrinsic shapes. We derived these slopes from the same calculation in equation \ref{eq:lin_fit}. The errors shown are the fitting errors of the slope.  These were calculated for the ITL sensor (R03-S12).}
    \label{fig:br}
\end{figure}

We would like to note here that the measurement of anisotropy in the BFE in the sensors could be affected by the intrinsic shape of the stars and their alignment on the pixel grid if the BFE is sensitive to the gradient of light between pixels. The stars are intrinsically shaped and oriented on the pixel grid as shown in Figure \ref{fig:spot_shapes} (based on their shapes at low signal level). Although there is little variation among the spots shapes in our sample, we attempt to measure how the BFE changes with respect to this intrinsic star shape in Figure \ref{fig:br}.  We measure the slope of the BFE individually for each star rather than the ensemble, as a function of intrinsic star shape, and we find that there is a small (but slightly beyond estimated error) decrease in the strength of the BFE in more extended sources. Since our stars are intrinsically larger in the y-direction than the x-direction, we would expect that $\alpha_{xx}$ in our sensors is slightly smaller than what we report and $\alpha_{yy}$ in our sensors is slightly larger than we report, therefore the overall anisotropy in our sensors is likely larger than what we report. However, given that there is only a small variation in intrinsic spot shapes, the error bars are on the same order of magnitude as our measurements in Figure \ref{fig:spot_shapes}, and it is impractical to extrapolate our measurements for perfectly round sources.

\subsection{Future Work}\label{ssec:bftocosmo}

The impact of BF residuals on future LSST science remains to be quantified, and steps can be taken to reduce the sensitivity of science to BFEs. BF can impact LSST weak lensing (WL) science by directly biasing survey targets, or indirectly by improper PSF estimation \citep{psfmodeling&gravlensing}. BF distortion effects on the bright stars in our study are of $O(1\%)$, which is larger than the $O(0.1-1\%)$ distribution of shape distortions caused by weak gravitational lensing (WL) in galaxy clusters, which is an observable sensitive to dark matter gravitational potentials along the line of sight and the dark energy equation of state \citep{Huterer2010,lsst_SciDoc}. 

%We use the \citet{Jarvis_2016} and \citet{heymans2006} formalism to estimate the error on the calibrated shear measurement of a galaxy that has been corrected with a biased PSF:

%\begin{equation}
%    \mathbf{e}_{\mathrm{meas}} = (1+m)\mathbf{e}_{\mathrm{true}} + \alpha \mathbf{e}_{\mathrm{PSF}} + \mathbf{\gamma},
%\end{equation}

%\noindent where $\mathbf{e} = [I_{xx} + I_{yy},\;I_{xx} - I_{yy}]$ is the galaxy shape or the PSF shape at the sensor location where the galaxy was measured, $m$ models the multiplicative bias due to the BFE on the sheared object itself, $\mathbf{\gamma}$ models additive biases, and $\alpha$ models the multiplicative bias on the PSF, which gets added to the bias on a single galaxy as a result of convolution.

\citet{lsst_SciDoc} science document requires us to reconstruct the true PSF size to a level of $\delta T_{PSF} / T_{PSF} < 10^{-3}$ in the co-added PSF in Y10, which in our lab-simulated stars is only satisfied below $2.5\times10^4\;\mathrm{el}$ in the ITL sensor and $2.3\times10^4\;\mathrm{el}$ in the E2V sensor, which covers only about 30\% of the dynamic range of these sensors. The current LSST science requirements do not specify an error budget for the $I_{xy}$ component.

The artificial stars in this study were nearly $2\times$ larger than the expected LSST PSF size (0.7" or 3.5px FWHM), and we would expect the BFE and the residual BFE to be worse for smaller PSFs due to the stronger charge gradients in the pixel plane. We also anticipate that bluer-bands will experience an augmented BFE due to a shorter photon conversion depth within the sensor, thereby resulting in a strengthened BFE from the increased drift length \citep[observed by][in HSC and ongoiong work to measure the effect in LSSTCam sensors]{Astier_2023}. In addition, higher-order radial moment errors in the PSF are known to also impact WL observables, which is also a subject of future investigation \citep{Zhang_Mandelbaum_Collaboration_2021, Zhang_Almoubayyed_Mandelbaum_Meyers_Jarvis_Kannawadi_Schmitz_Guinot_Collaboration_2022}. 

Luckily, the BF correction benefits with the incorporation of higher order BFEs into current correction schemes, and informed star selection from improved sensor characterization.

\subsection{Proposed Improvements}

The correction of LSST PSF-like stars is good enough for LSST weak lensing science requirements below a certain signal level, and this helps to inform the selection of PSF stars to estimate the PSF. We will therefore need to apply the methods in this paper to all LSSTCam sensors since different sensor types and sensors can have different characteristics overall. 

In addition, there are several steps that could be added and tested to the LSST Science Pipelines that could improve the correction:

\begin{enumerate}
    
    \item In the scalar model defined by \citet{Coulton_2018}, change the form of equation \ref{bfcorrectioneqn} ($\hat{F} = F + \frac{1}{2} \nabla \cdot V $, where $V = F\nabla (K\otimes F)$) to enforce Gauss's law on small scales such that the charge lost over the area of one pixel is exactly matched by the sum of the charges lost over each boundary ($X$) of the pixel (such that $\int \nabla \cdot V dx^2 = \Sigma_X V_X$). This will ensure charge conservation and improve the local modeling of charge transport due to the BFE (Lance Miller, \href{mailto:Lance.Miller@physics.ox.ac.uk}{Lance.Miller@physics.ox.ac.uk}, in private communication). If the lack of charge conservation originates from the small scale deviations from physics, then this implementation should fully conserve charge.

    \item Include a curl component of the deflection field. This could be done by using a electrostatically derived vector-based model fit to the pixel-pixel correlations in flat fields, with each pixel boundary shift being a free parameter. Such a method is proposed in \citet{Astier_2023} and tested for HSC, but it should be tested in LSSTCam. 

    \item Improve the NL correction at higher signal levels in LSSTCam. The best-fit model parameters of the PTC are somewhat sensitive to the residual NL for these sensors since we fit up to the pCTI turnoff, which is above the signal level we start to observe significant signal NL. In addition, the PTC scanning method we describe is sensitive to the signal range on which we calculate the $\chi^2$ for the same reason. While we mitigate this by limiting the signal range, the method we propose here does require the adequate removal of NL and other sources of pixel correlations, or it at least it requires plenty of flat-pairs at low enough signal levels that one can still accurately characterize the BFE without the impact of these other effects at high signal levels.

\end{enumerate}

\section{Conclusions}
In this paper we measured the BFE in two types of sensors in the LSSTCam focal plane from flat-field and artificial stars in the lab. We then quantified and qualified the scalar BF correction proposed by \citet{Coulton_2018} with improvements to account for sources of higher-order pixel correlations. 

First, we measured dense flat pair data to characterize the PTC, and we found that the BFE gets stronger with signal due to higher-order, non-linear BFE components in a model by \citet{Astier_2019}, which includes flux-dependent drift fields and feedback mechanisms on accumulating charges. The BF correction proposed in \citet{Coulton_2018}, which uses a scalar kernel derived from the pixel covariance matrix, relies on the assumption of linearity. The merit of the correction therefore depends on the signal level from which we calculate the covariance matrix and derive the kernel, which is a priori ambiguous.

Secondly, we introduced a new procedure to derive the kernel in a way that includes the contribution of non-linear BFE terms in the correction. We determined the kernel that best reconstructs the linear PTC. We then used this kernel to correct ithe BFE n both sensor types by 95-97\%, but was only able to correct up to a flux we identify as the parallel deferred charge turnoff at $10^5$ $\mathrm{el}$, where we can no longer efficiently transfer charge along an image column during readout. The flux range of pixels that can be used for weak lensing science is severely limited by this cutoff.

Thirdly, we corrected the BFE in artificially produced stars by $96-99\%$ in the serial direction and by $85-90\%$ in the parallel direction. The BFE was observed to be anisotropic (16\%) relative to the CCD coordinate system, and the correction preserved 67\% of this anisotropy.  

We eliminated other systematics that could have interfered with these results, including non-circular artificial stars and charge-conservation, to identify the source of these effects as intrinsic BFEs in the sensor itself and limitations of the assumptions of the \citet{Coulton_2018} correction. Our data suggests that the higher-order BFEs are not negligible and that the zero-curl assumption of the \citet{Coulton_2018} algorithm is not fully valid.

Our findings also motivate a detailed study on more realistic PSF stars and how measurement errors from BF could ultimately impact cosmology and other science goals. Ultimately, it is important to characterize the sensitivity of cosmological parameters to observables biased with BFEs. Even with state-of-the-art correction techniques, the residual effects could represent a significant component of the systematics error budget for cosmological analyses of LSST observations.

%One can also consider the impact that the BFE has on Type 1a supernova (SN) photometry and other transient searches. Supernovae are known to be bright, often outshining their host galaxies. LSST is expected to be capable of probing unlensed Type Ia SNe down to about $22\mathrm{nd}$ \textit{r}-band magnitude at a redshift of $z=1.3$, with a goal of $\mathrm{mmag}$ level photometric precision ($0.1\%$), and it will be a direct measurement of the Hubble parameter. Photometric errors due to BF or spontaneous charge loss from previous correction techniques could directly propagate into distance errors or maximum likelihood estimators of SNe identification. Further study is required to understand the impact of unmodeled BFE components on on past and future SN identification and photometry.

\newpage
\section*{Acknowledgements}
This material is also based upon work supported in part by the National Science Foundation through Cooperative Agreement AST-1258333 and Cooperative Support Agreement AST-1202910 managed by the Association of Universities for Research in Astronomy (AURA), and the Department of Energy under Contract No. DE-AC02-76SF00515 with the SLAC National Accelerator Laboratory managed by Stanford University. Additional Rubin Observatory funding comes from private donations, grants to universities, and in-kind support from LSSTC Institutional Members. The work of AB and SM is also supported by the Department of Energy (DOE) grant DE-SC0009920. The work of AB was partially supported by the DOE Office of Science Graduate Student Research (SCGSR) Award. We are also grateful for the expert help and advice from Chris Waters (\href{mailto:czw@princeton.edu}{czw@princeton.edu}) for his expertise and ingenuity with the LSST Science Pipelines. We also acknowledge the help of Lance Miller (\href{mailto:Lance.Miller@physics.ox.ac.uk}{Lance.Miller@physics.ox.ac.uk}) on behalf of the Euclid consortium for his work identifying and implementing the aforementioned flux-conserving BF correction into the LSST Science Pipelines, which is currently under development. In addition, we would like to acknowledge the helpful wisdom and guidance of Pierre Astier (\href{mailto:pierre.astier@in2p3.fr}{pierre.astier@in2p3.fr}) and Pierre Antilogus (\href{mailto:pierre.antilogus@in2p3.fr}{pierre.antilogus@in2p3.fr}).

\software{LSST Science Pipelines (\url{https://pipelines.lsst.io}), MixCOATL (\url{https://github.com/Snyder005/mixcoatl}), GalSim (\url{https://github.com/GalSim-developers/GalSim})}

\bibliography{refs}{}
\bibliographystyle{aasjournal}

\appendix

\section{Instrument Signature Removal Configurations}\label{appendix:configs}

In Table \ref{tbl:isr_configs} we present the configurations we used with the LSST Science Pipelines for ISR of the spot images \citep{LsstDMPipeline1,LsstDMPipeline2}. For spot identification and characterization, we use an LSST Science Pipeline wrapper called the Mixed Calibration Optics Analysis Test Library (MixCOATL). Our data processing also utilized code (with default configurations) for matching and labeling spots to specific holes in the lithographic mask of the optical projector (\texttt{GridFitTask}). This code is described in \citet{JohnnyPaper} and integrated into the LSST Science Pipelines. For each task, including the calibration and ISR tasks, we used the default configuration parameters.

\begin{table*}[t]
    \centering
    Configurations for Instrument Signature Removal
    \begin{tabular*}{\textwidth}{@{\extracolsep{\fill}}lc}
        \toprule[1pt] % <-- Toprule here
        \midrule[0.3pt]
        Parameter & Value \\
        \midrule[0.3pt]
        \texttt{config.doSaturation} & False \\
        \midrule[0.3pt]
        \texttt{config.growSaturationFootprintSize} & 0 \\
        \midrule[0.3pt]
        \texttt{config.doSuspect} & False \\
        \midrule[0.3pt]
        \texttt{config.edgeMaskLevel} & 'DETECTOR' \\
        \midrule[0.3pt]
        \texttt{config.doOverscan} & True \\
        \midrule[0.3pt]
        \texttt{config.overscan.fitType} & 'MEDIAN' \\
        \midrule[0.3pt]
        \texttt{config.overscan.order} & 1 \\
        \midrule[0.3pt]
        \texttt{config.overscan.numSigmaClip} & 3.0 \\
        \midrule[0.3pt]
        \texttt{config.overscan.maskPlanes} & ['BAD', 'SAT'] \\
        \midrule[0.3pt]
        \texttt{config.overscan.overscanIsInt} & True \\
        \midrule[0.3pt]
        \texttt{config.overscan.doParallelOverscan} & False \\
        \midrule[0.3pt]
        \texttt{config.overscan.parallelOverscanMaskThreshold} & 100000 \\
        \midrule[0.3pt]
        \texttt{config.overscan.parallelOverscanMaskGrowSize} & 7 \\
        \midrule[0.3pt]
        \texttt{config.doBias} & True \\
        \midrule[0.3pt]
        \texttt{config.doBiasBeforeOverscan} & False \\
        \midrule[0.3pt]
        \texttt{config.doDeferredCharge} & True \\
        \midrule[0.3pt]
        \texttt{config.deferredChargeCorrection.nPixelOffsetCorrection} & 15 \\
        \midrule[0.3pt]
        \texttt{config.deferredChargeCorrection.nPixelTrapCorrection} & 6 \\
        \midrule[0.3pt]
        \texttt{config.deferredChargeCorrection.useGains} & False \\
        \midrule[0.3pt]
        \texttt{config.deferredChargeCorrection.zeroUnusedPixels} & False \\
        \midrule[0.3pt]
        \texttt{config.doLinearize} & True \\
        \midrule[0.3pt]
        \texttt{config.doCrosstalk} & False \\
        \midrule[0.3pt]
        \texttt{config.doDefect} & True \\
        \midrule[0.3pt]
        \texttt{config.doNanMasking} & True \\
        \midrule[0.3pt]
        \texttt{config.doWidenSaturationTrails} & True \\
        \midrule[0.3pt]
        \texttt{config.doBrighterFatter} & True \\
        \midrule[0.3pt]
        \texttt{config.doFluxConservingBrighterFatterCorrection} & True \\
        \midrule[0.3pt]
        \texttt{config.brighterFatterLevel} & 'DETECTOR' \\
        \midrule[0.3pt]
        \texttt{config.brighterFatterMaxIter} & 10 \\
        \midrule[0.3pt]
        \texttt{config.brighterFatterThreshold} & 10.0 \\
        \midrule[0.3pt]
        \texttt{config.brighterFatterApplyGain} & True \\
        \midrule[0.3pt]
        \texttt{config.brighterFatterMaskGrowSize} & 0 \\
        \midrule[0.3pt]
        \texttt{config.doDark} & True \\
        \midrule[0.3pt]
    \end{tabular*}
    \caption{The configuration parameters for \texttt{lsst.ip.isr.isrTask.IsrTask} in the LSST Science Pipelines \citep{LsstDMPipeline1,LsstDMPipeline2}.}\label{tbl:isr_configs}
\end{table*}
\newpage
\begin{table*}[t!] \ContinuedFloat
    \centering
    Configurations for Instrument Signature Removal
    \begin{tabular*}{\textwidth}{@{\extracolsep{\fill}}lc}
        \toprule[1pt] % <-- Toprule here
        \midrule[0.3pt]
        Parameter & Value \\
        \midrule[0.3pt]
        \texttt{config.doStrayLight} & False \\
        \midrule[0.3pt]
        \texttt{config.doFlat} & False \\
        \midrule[0.3pt]
        \texttt{config.doApplyGains} & False \\
        \midrule[0.3pt]
        \texttt{config.usePtcGains} & True \\
        \midrule[0.3pt]
        \texttt{config.doFringe} & False \\
        \midrule[0.3pt]
        \texttt{config.doAmpOffset} & False \\
        \midrule[0.3pt]
        \texttt{config.doInterpolate} & False \\
        \midrule[0.3pt]
        \texttt{config.doSaturationInterpolation} & False \\
        \midrule[0.3pt]
        \texttt{config.doNanInterpolation} & True \\
        \midrule[0.3pt]
        \texttt{config.doVignette} & False \\
      \bottomrule[0.3pt] % <-- Bottomrule here
    \end{tabular*}
    \caption{ Continued.}
\end{table*}

After this sequence of ISR, we modeled and subtracted the background light produced by the spot grid projector. Given that the background light produces photocharges that get collected in the potential wells of our sensor, it contributes to the BFE, and should be taken into account when applying the scalar kernel and correcting an image. However, this component should be removed before shape-fitting. We performed this background subtraction separately for both the corrected and uncorrected images before calculating and cataloging the final shape statistics. We performed this using a $65 \times 65$px median filter to match the artificial star spacing, masking over the fitted footprints of the artificial stars.

\newpage

\end{document}